\documentclass[a4paper,11pt]{article}
\usepackage{etex}
\usepackage[utf8]{inputenc}
\usepackage{mathtools,amsmath,amssymb}
\usepackage{graphicx}
\usepackage{lmodern}
\usepackage[T1]{fontenc}
\usepackage{microtype}
\usepackage[pagebackref=false]{hyperref}
\numberwithin{equation}{section}

\usepackage{xspace}
\usepackage{bibentry}

\usepackage{graphicx}

\usepackage{subcaption}

\usepackage[numbers,sort&compress]{natbib}
\usepackage{hypernat}

\usepackage{tikz}
\usetikzlibrary{decorations.pathreplacing}
\usepackage[margin=2.5cm]{geometry}

\definecolor{todocolor}{HTML}{D7E1E5}
\usepackage[textsize=tiny, 
            backgroundcolor=todocolor, 
            bordercolor=todocolor, 
            linecolor=todocolor, 
            textwidth=3.9cm]{todonotes}

\usepackage{comment}
\usepackage{rotating}

\def\be{\begin{equation}}
\def\ee{\end{equation}}
\def\bea{\begin{eqnarray}}
\def\eea{\end{eqnarray}}

\def\<{\langle}
\def\>{\rangle}
\def\~{\tilde}

\def\l{\lambda}

\newcommand{\N}{\Bbb N}

\newcommand{\Z}{\Bbb Z}

\newcommand{\beq}{\begin{eqnarray}}
\newcommand{\eeq}{\end{eqnarray}}

\newcommand{\ID}{I}

\DeclareMathOperator{\tr}{tr}

\DeclareMathOperator{\diag}{diag}

\newcommand{\KK}{\mathbb{S}}
\newcommand{\kk}{s}

\newcommand{\de}{\delta}

\newcommand{\diagram}{
\tikzstyle{redRectangle} = [
    rectangle,
    draw,
    fill=red!20,
    node distance=2 cm,
    text width=7 em,
    text centered,
    rounded corners,
    minimum height=4 em,
    minimum width=3 cm,
    thick
]

\tikzstyle{blueRectangle} = [
    rectangle,
    draw,
    fill=blue!20,
    node distance=2 cm,
    text width=7 em,
    text centered,
    rounded corners,
    minimum height=4 em,
    minimum width=3 cm,
    thick
]

\tikzstyle{yellowRectangle} = [
    rectangle,
    draw,
    fill=yellow!20,
    node distance=2 cm,
    text width=7 em,
    text centered,
    rounded corners,
    minimum height=4 em,
    minimum width=3 cm,
    thick
]

\tikzstyle{greenRectangle} = [
    rectangle,
    draw,
    fill=green!20,
    node distance=2 cm,
    text width=7 em,
    text centered,
    rounded corners,
    minimum height=4 em,
    minimum width=3 cm,
    thick
]

\tikzstyle{empty} = [
]

\tikzstyle{line} = [
    draw,
->,
    thick
]

\begin{tikzpicture}

    \node [empty](origin){};
    
    \node [greenRectangle, above=of origin](q0){Povolotsky's model \cite{Povolotsky}};
    \node [greenRectangle, right=1.5cm of q0] (q1) {q-Hahn AZRP \cite{barraquand}};
    \node [greenRectangle, below=1.5cm of q1] (qsc1) {MADM \cite{Sasamoto}};
    \node [yellowRectangle, right=of q1](q2){non-compact XXX spin $s$};
    \node [yellowRectangle, right=of qsc1] (qsc2) {non-compact XXX spin $\frac{1}{2}$};

    \path [line] (q1) -- (q2);
    \path [line] (qsc1) --  (qsc2);
    \path [line] (q0) -- (q1);
    \path [line] (q1) --  (qsc1);
    \path [line] (q2) --  (qsc2);

\end{tikzpicture}

}

\begin{document}

\begingroup\parindent0pt
\centering
\begingroup\LARGE
\bf
Non-compact quantum spin chains as \\ integrable stochastic particle processes
\par\endgroup
\vspace{3.5em}
\begingroup\large
{\bf Rouven Frassek}$\,^a$, 
{\bf Cristian Giardin\`{a}}$\,^b$, 
{\bf Jorge Kurchan}$\,^c$ 
\par\endgroup
\vspace{2em}
\begingroup\sffamily\footnotesize
$^a\,$Max-Planck-Institut für Mathematik,\\
   Vivatsgasse 7, 53111 Bonn, Germany\\
\vspace{1em}
$^b\,$University of Modena and Reggio Emilia, FIM,\\
Via G. Campi 213/b, 41125 Modena, Italy\\
\vspace{1em}
$^c\,$Laboratoire de Physique de l’Ecole Normale Supérieure, ENS,\\ Université PSL, CNRS, Sorbonne Université,\\ Université Paris-Diderot, Sorbonne Paris Cité, Paris, France\\
\par\endgroup
\vspace{5em}


To Joel Lebowitz, for his continuous inspiration.

\vspace{1.cm}
\begin{abstract}
\noindent
In this paper we discuss a family of models of particle and energy diffusion
on a one-dimensional lattice, related to those studied previously in 
\cite{Sasamoto}, \cite{barraquand} and \cite{Povolotsky}
in the context of KPZ universality class.
We show that
they may be mapped onto  an integrable $\mathfrak{sl}(2)$
Heisenberg spin chain whose Hamiltonian density in the bulk has been already 
studied in the AdS/CFT and the integrable system literature. 

Using the quantum inverse scattering method, we study various new aspects, in particular we identify 
boundary terms, modeling reservoirs in non-equilibrium statistical mechanics models, for which the spin chain 
(and thus also the stochastic process) continues to be integrable. 
We also show how the construction of a "dual model" of probability theory is possible 
and useful.

The fluctuating hydrodynamics  of our stochastic
model corresponds to the semiclassical evolution of a string that derives from 
 correlation functions of local gauge invariant operators of $\mathcal{N}=4$ super Yang-Mills theory (SYM), in imaginary-time. As any stochastic system, it has a supersymmetric completion that encodes
for the thermal equilibrium theorems: we show that in this case it is equivalent to the $\mathfrak{sl}(2|1)$ superstring
that has been derived directly  from $\mathcal{N}=4$ SYM.

\end{abstract}

\endgroup

\thispagestyle{empty}
\setcounter{tocdepth}{2}

\newpage
\tableofcontents
\newpage

\section{Introduction }

\subsection{The setting}
Stochastic systems may be described by a linear operator -- called here the Hamiltonian $\cal{H}$ -- 
that describes the infinitesimal evolution of the probability distribution. This operator has
to guarantee probability conservation, and obviously be such that transition rates are real and positive.
In many cases, such as the overdamped Langevin equation, $\cal{H}$ is, when written in some appropriate base, Hermitean. When this happens,
we have a direct connection between the original stochastic system and a quantum one: the former evolves with Euclidean time ( $e^{-t \cal{H}}$) and 
the latter with "real" time ($e^{-i t \cal{H}}$).
In particular, if we compute the sum over periodic trajectories of period $\beta$ of the stochastic model, we are in fact
calculating the partition function of the "quantum" system with temperature $1/\beta$.
Establishing a link between a stochastic and a quantum model is illuminating, and allows to exchange techniques between two different fields.
It sometimes happens that a {\em single} Hamiltonian has more than one basis in which it may be interpreted as a stochastic process. 
The  picture described by these {\em dual} models may be quite different: for example  one may describe the transport of a continuous quantity,
interpreted as energy, while its dual the transport of discrete particles. We shall meet this situation here.

\smallskip
An example in point is the Symmetric Exclusion Process (SEP) in which particles move randomly to the right or to the left of a
one-dimensional grid but are not
allowed to superpose. It is well known that the evolution operator may be mapped onto a one-dimensional ferromagnetic chain 
${\cal H} = \sum_i \left[ 2\sigma_0^{[i]} \sigma_0^{[i+1]} + \sigma_+^{[i]} \sigma_-^{[i+1]} + \sigma_-^{[i]} \sigma_+^{[i+1]}\right]$ of spins
one-half, satisfying the $\mathfrak{su}(2)$ algebra: 
\begin{equation}\label{eq:su2com}
 [\sigma_0,\sigma_\pm]=\pm \sigma_\pm,\qquad [\sigma_+,\sigma_-]=2\sigma_0 \,. 
\end{equation} 
Another popular system is the Kipnis-Marchioro-Presutti (KMP) model \cite{kipnis1982heat}, where each pair of neighboring sites exchange randomly their energies. 
It was realized years ago \cite{giardina2007duality, GKR}  that it may be put in direct relation with a chain of 
  $\mathfrak{sl}(2)$ `spins'  ${\cal H} = \sum_i \left[ 2S_0^{[i]} S_0^{[i+1]} - S_+^{[i]} S_-^{[i+1]} -S_-^{[i]} S_+^{[i+1]}\right]$ satisfying  at each site of the spin chain  
 the commutation relations
\begin{equation}\label{eq:sl2com}
 [S_0,S_\pm]=\pm S_\pm,\qquad [S_+,S_-]=-2S_0 \,. 
\end{equation} 
This mapping immediately explained why many formulas for the KMP model could be obtained directly as an analytic continuation of the SEP
ones: the $\mathfrak{sl}(2)$ algebra may be though of as "negative spin" representations of the $\mathfrak{su}(2)$ algebra.
The relation  has however a disappointing side: although the spin one-half $\mathfrak{su}(2)$ chain is integrable \cite{Bethe1931}, 
the $\mathfrak{sl}(2)$ one associated with KMP is not. 

\smallskip
Non-compact {\em integrable}  spin chains were studied in theoretical physics, in relation to high energy
QCD \cite{Lipatov:1993yb,Faddeev:1994zg,Braun:1998id}, $\mathcal{N}=4$ super Yang-Mills theory ($\mathcal{N}=4$ SYM) \cite{Minahan:2002ve,Beisert:2003yb,Beisert:2003jj} and the AdS/CFT dual string theory limit considered in \cite{kruczenski,bellucci-sl2,Stefanski:2004cw}. However, so far their interpretation as stochastic processes has not been studied. 
In this paper we address precisely this question.

\subsection{Models and relation to previous literature}

We shall be  interested in the relation between integrable 
non-compact spin chains and interacting particle systems, 
especially in the context of {\em non-equilibrium statistical mechanics}. 
Thus we shall consider the role of boundary stochastic processes, which
amounts to including boundary terms in the spin chains. For the particle/energy process, the 
boundary processes model infinite reservoirs that fix the chemical potentials/temperatures at the
boundaries. When their values are different, a non-equilibrium steady state sets in, with a non-zero current. 
It turns out that the class of {\em boundary-driven models} that we introduce have a bulk
part that coincides with the {\em symmetric} version of several particle processes that
have been previously studied on the infinite line $\mathbb{Z}$ as microscopic models for 
the Kardar-Parisi-Zhang (KPZ) universality class \cite{kardar1986dynamic, sasamoto2010one, corwin2012kardar}.

\smallskip
We start with the case of spin $s=\frac{1}{2}$ for which 
we introduce an {\em open} integrable Hamiltonian (Section
\ref{spin12}). The bulk part of this Hamiltonian was first studied by 
N.~Beisert in \cite{Beisert:2003jj}; the closed spin chain 
also appears in weakly-coupled $\mathcal{N}=4$ super 
Yang-Mills theory in the so-called $\mathfrak{sl}(2)$ sector.
The boundary interaction we consider is novel and, remarkably, 
preserves integrability.
Seen as a stochastic operator, the open Hamiltonian 
is -- in an appropriate base choice -- 
the probability evolution operator for an interacting particle 
system. We show (Section \ref{sasa}) that this process happens to be 
the boundary-driven version of the Multi-particle Asymmetric 
Diffusion Model (MADM) of Sasamoto-Wadati \cite{Sasamoto}. The equivalent drop-push model has been studied in \cite{alimohammadi1998exact,alimohammadi1999two}.

\smallskip
Furthermore, again for spin $s=\frac{1}{2}$, we show that
from the open Hamiltonian one can also get boundary driven
processes taking values in the continuum, that can be used
to model energy transport. This is achieved by taking a scaling 
limit that leads to integral operators. In this way one obtains  
a Markov process in the form of a L\'evy process (Section \ref{scaling}).
For this process we have not been able to find a reference in 
the literature. The bulk part is however similar, yet different, 
from the expression given in \cite{Derkachov:1999pz}, which
indeed is not stochastic.

\smallskip
We also discuss briefly the case of general spin $s$  (Section \ref{povo}).
The integrable Hamiltonian of the higher spin models can be obtained 
from the expression in terms of integral operators \cite{Derkachov:1999pz}. 
Seen as a stochastic operator the bulk Hamiltonian density
is related to the q-Hahn Asymmetric Zero Range Process  (AZRP) of 
Barraquand and Corwin \cite{barraquand}, which in turn generalizes
the q-Hahn Totally Asymmetric Zero Range Process (TAZRP) introduced 
in \cite{Povolotsky} by allowing jumps in both directions. 
Algebraically, these models and their multi species 
generalizations can be described using the stochastic 
R-matrix \cite{Kuniba}. In contrast to the stochastic R-matrix approach,
where the particle process is described in terms of two commuting Hamiltonians 
which generate left and right moving particles separately, we find that the standard 
nearest-neighbor Hamiltonian of the non-compact spin chain yields immediately a 
process of particles hopping to the left and the right. 
The discrete-time infinite-lattice version of the proposed XXX dynamics appeared 
as random walks in random environment in \cite{barraquand2017random}.
Also different rational limits of the q-Hahn process including the ones beyond the 
stochastic sector where discussed in context of directed polymers in
random environment, see e.g. \cite{thiery2015integrable}.

\begin{figure}
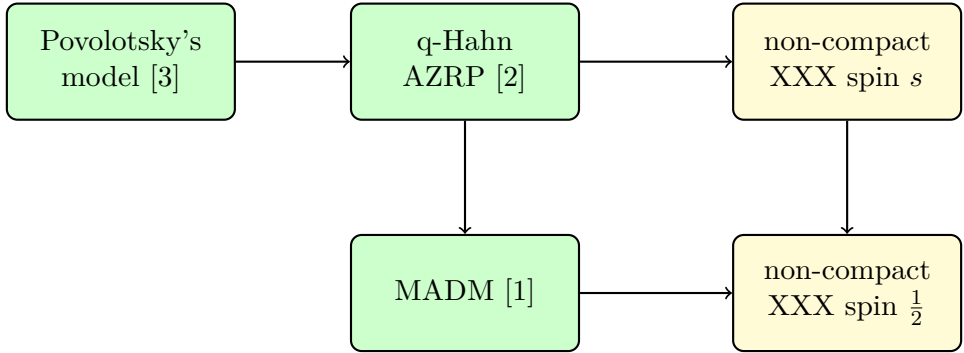

\begin{center}  
\diagram
\end{center}  
\caption{Relation between the stochastic hopping models in green boxes and  the spin chains in yellow boxes. 
The arrows indicate that the process can be obtained as a limit. }
\label{fig:graph}
\end{figure}

\smallskip
We summarize the connections between integrable interacting
particle systems and integrable non-compact spin chains in 
Figure \ref{fig:graph}.

\subsection{Informal description of the main results}

\smallskip

For the benefit of the reader we summarize here the main results of this paper:

\medskip
{\bf (i)}  We study  continuous time stochastic 
processes that arise from rational non-compact spin chains. 
More precisely we introduce a class of boundary-driven 
processes that maps to the integrable non-compact {\em open} Heisenberg 
XXX spin chains. The Hamiltonian is thus made by a {\em bulk part}, which is
the sum of nearest neighbors terms, and by a {\em boundary part},
which involves only the first and last spin of the chain. 
We identify boundary terms that preserve integrability.

\medskip
{\bf (ii)}
It is well known that integrable non-compact spin chains can be described in the framework of the {\em quantum inverse scattering method}. This allows to determine the family of commuting operators in terms of transfer matrices and gives access to the spectrum, eigenvectors and observables via the algebraic Bethe ansatz, separation of variables or functional methods. Further, knowing the algebraic structure of the bulk model one can construct integrable boundary models following Sklyanin~\cite{Sklyanin:1988yz}. We exemplify this procedure in the case of the rational non-compact spin~$\frac{1}{2}$ $\mathfrak{sl}(2)$ chain (Section \ref{sec:qism}).
We deduce certain integrable stochastic boundary conditions from the most general K-matrix that we derive from the boundary Yang-Baxter equation. Furthermore we  comment on the application of the algebraic Bethe ansatz to this model.

\medskip
{\bf (iii)} 
We show that our boundary-driven systems admits a {\em dual
process} (Section \ref{sec:duality}).
Again, for a proof we restrict to the particle process of spin $\frac{1}{2}$,
however the result is more general and in particular it applies
to general spin $s$, as well as to the Levy process obtained
in the scaling limit. The dual process has two absorbing states,
thus the non-equilibrium steady state of the original
system is encoded in the {\em absorption probabilities of
dual particles}. We illustrate this by analyzing the $n$-point
correlation functions in the stationary state.

\medskip
{\bf (iv)} We  investigate the connection of our models with $\mathcal{N}=4$ super Yang-Mills theory in the setting of a closed chain. 
The limit of {\em fluctuating hydrodynamics} of the $\mathfrak{sl}(2)$  chain, describing 
coarse-grained properties of the stochastic system, turns out to be nothing but the semiclassical string equation,  in Euclidean time.   
Indeed, both derivations of infrared properties  \cite{kruczenski,bellucci-sl2,Tailleur} are virtually identical, and have been 
independently made with coherent states  (in the stochastic case more rigorous constructions of the theory of large deviations 
around hydrodynamic limit exist \cite{jona}). One may ask why a stochastic model would appear in the SYM context: part of the 
answer is probably the underlying supersymmetry. We describe this in Section \ref{sec:4sym}, where we shall also show how 
{\em supersymmetry} arises in this stochastic model.

\section{Non-compact spin chains as integrable stochastic process}\label{sec:model}

\subsection{The case of spin $s=1/2$}
\label{spin12}
In the following we study aspects of the non-compact $\mathfrak{sl}(2)$ spin $\frac{1}{2}$ Heisenberg spin chain in the setting of open boundaries. The nearest neighbor Hamiltonian of the model is of the form
\begin{equation}\label{eq:fullham}
\mathcal{H}=\mathcal{H}_1+\sum_{i=1}^{N-1}\mathcal{H}_{i,i+1}+\mathcal{H}_N\,.
\end{equation} 
We focus on spin chains with infinite-dimensional discrete series representation with lowest weight such 
that the representation at each site of the spin chain is given via
\begin{equation}\label{eq:sl2action}
 S_+|m\rangle =(m+1)|m+1\rangle\,,\quad  S_-|m\rangle =m|m-1\rangle\,,\quad  S_0|m\rangle =(m+\frac{1}{2})|m\rangle\,,
\end{equation} 
Here the generators of the $\mathfrak{sl}(2)$ algebra at each site of the spin chain satisfy the commutation relations \eqref{eq:sl2com}
and $|0\rangle$ denotes the lowest weight state.
Then the action of the Hamiltonian \eqref{eq:fullham} on quantum space $V$ of the spin chain 
\begin{equation}
 V=|m_1\rangle \otimes\ldots \otimes|m_N\rangle\,,
\end{equation} 
is defined as follows.
The Hamiltonian density $ \mathcal{H}_{i,i+1}$  for the spin $\frac{1}{2}$ chain acts on two sites and can be written as
\begin{equation}\label{eq:harmonicaction}
\begin{split}
 \mathcal{H}_{i,i+1}|m_i\rangle\otimes|m_{i+1}\rangle=\left(h(m_i)+h(m_{i+1})\right)|m_i\rangle\otimes|m_{i+1}\rangle
 &-\sum_{k=1}^{m_i}\frac{1}{k}|m_i-k\rangle\otimes|m_{i+1}+k\rangle\\&-\sum_{k=1}^{m_{i+1}}\frac{1}{k}|m_i+k\rangle\otimes|m_{i+1}-k\rangle,
 \end{split}
\end{equation}  
with the harmonic numbers 
\begin{equation}\label{eq:harmonicnumbers}
 h(m)=\sum_{k=1}^m\frac{1}{k}\,.
\end{equation} 
This particular form of the Hamiltonian was studied in the context of $\mathcal{N}=4$ super Yang-Mills theory in  \cite{Beisert:2003jj}.
The closed chain is known to be integrable and can be written in the standard form, cf.~\cite{Faddeev:1996iy}, in terms of the digamma function $\psi(x)$ defined as the logarithmic derivative of the Gamma function $\Gamma(x)$ as
\begin{equation}
\label{ham:fadeev}
 \mathcal{H}_{i,i+1}=2(\psi(\KK_{i,i+1})-\psi(1))\,.
\end{equation} 
Here, the operator $\KK_{i,i+1}$ is related to the two-site Casimir acting on two sites $i$ and $i+1$ via
$C_{i,i+1}=(S^{[i]}+ S^{[i+1]})^2= \KK_{i,i+1}(\KK_{i,i+1}-1)$. The operator $\mathcal{H}$ is defined by its action on the eigenbasis of $\mathbb{S}$. 
\footnote{
A formula for the Hamiltonian density in terms of the the two site Casimir $C_{i,i+1}$ has been discussed in \cite{Faddeev:1996iy,Korchemsky:1994um}.}
The eigenvalues of $\KK$ are determined from the irreducible tensor product decomposition
\begin{equation}\label{eq:irreps12}
\Big[\frac12\Big]\otimes \Big[\frac12\Big]=\bigoplus_{j=0}^\infty\Big[\Lambda_j\Big],
\end{equation} 
where $\Big[\frac12\Big]$ denotes the representation of spin $\frac12$ as defined by the module in \eqref{eq:sl2action} and $[\Lambda_j]$ the corresponding representation labeled by the spin value $\Lambda_j=1+j$, i.e. the module defined in \eqref{eq:sl2spinsaction} when replacing $s\to1+j$. 
This is in analogy to the selection rules for the addition of ordinary spin angular momentum. 
The action of the operator $\KK$ is diagonal on each term on the right hand side of \eqref{eq:irreps12}. Its eigenvalues are degenerate due to the $\mathfrak{sl}(2)$ invariance and simply given by $\Lambda_j$. Thus \eqref{ham:fadeev} immediately yields the eigenvalues of the harmonic action \eqref{eq:harmonicaction}.

In addition to the bulk terms we introduce the integrable boundary terms 
\begin{equation}\label{eq:harmbnd}
  \mathcal{H}_i|m_i\rangle=\left(h(m_i)+\sum_{k=1}^\infty\frac{\beta_i^k}{k}\right)|m_i\rangle -\sum_{k=1}^{m_i} \frac{1}{k}|m_i-k\rangle -\sum_{k=1}^\infty \frac{\beta_i^k}{k}|m_i+k\rangle ,
\end{equation} 
where $i=1,N$ and $0<\beta_1,\beta_N<1$. To our knowledge these types of boundary conditions have not been considered so far.

\begin{figure}
  \begin{center}  
 \includegraphics[width=1\columnwidth]{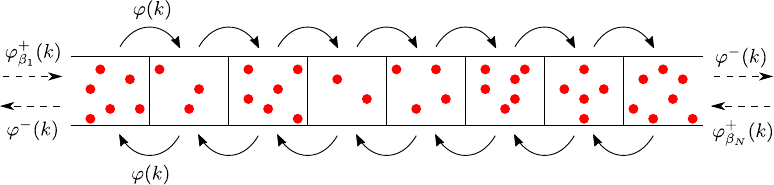}
  \end{center}  
  \caption{Stochastic hopping model emerging from the non-compact spin $\frac{1}{2}$ Heisenberg spin chain for the case of length $N=8$. Here the $i$th site corresponds to the $i$th box. The rates $\varphi(k)$ with $i=1,\ldots,7$ are determined from the integrable nearest-neighbor bulk Hamiltonian while $\varphi^\pm(k)$ emerge from the boundary terms.}
    \label{fig:gaz}
\end{figure}

The Hamiltonian \eqref{eq:fullham} defined above is stochastic because its matrix elements
outside of the diagonal are non-positive and the sum
over its columns vanishes. In other words, $-\mathcal{H}^t$ (i.e. the negative of the transposed of the Hamiltonian)
is the generator of a continuous time Markov process $\{{\bf M}(t), \, t\ge 0\}$,
representing a {\em system of interacting particles}.
Here ${\bf M}(t) = (M_1(t),M_2(t), \ldots, M_N(t))$ and
the $i^{th}$ component $M_i(t)$ denotes the number
of particles at site $i \in \{1,\ldots,N\}$ at time $t\ge 0$.
In the bulk particles jumps symmetrically to their 
nearest neighbour sites, so that $k\in \mathbb{N}$ particles (if available) moves
to the left or to the right a rate $\varphi(k)=1/k$.
At the boundary site ``$1$'', $k\in \mathbb{N}$ particles are created at rate $\varphi^+_{\beta_1}(k)={\beta_1^k}/{k}$,
where $\beta_1 \in (0,1)$ is parameter,
and $k \in \mathbb{N}$ particles (if available) are removed at a rate $\varphi^-(k)=1/k$.
A similar process, now with a parameter $\beta_N \in (0,1)$, occurs at the boundary site $N$.
See Figure \ref{fig:gaz} for a pictorial representation.

One can easily show that the particular case $\beta_1 = \beta_N = \beta$
describes the {\em equilibrium} set-up with a Boltzmann-Gibbs invariant measure.
More precisely the product measure with marginal the Geometric distribution
of parameter $0<\beta<1$,
i.e. the law
\be
\label{geo}
\mathbb{P}({\bf M}(t) = {\bf m}) = \prod_{i=1}^N \beta^{m_i} (1-\beta), 
\ee
where ${\bf m} = (m_1,\ldots,m_N) \in \N_0^N$ denotes a particle configuration, 
is reversible (and thus stationary) for the Markov process 
defined by the Hamiltonian  \eqref{eq:fullham}.
The generic case $\beta_1 \neq  \beta_N$ yields instead
a boundary driven  {\em non-equilibrium} interacting particle system,
whose invariant measure is unknown.

\subsection{Rational limit of Sasamoto\,-Wadati model}
\label{sasa}
The Multiparticle Asymmetric Diffusion Model (MADM) was introduced in \cite{Sasamoto} by Sasamoto and Wadati.
In the following we show that the bulk part of the Hamiltonian of the non-compact spin $\frac{1}{2}$ in \eqref{eq:harmonicaction}  can be obtained 
from the rational limit of the MADM, cf. \cite{alimohammadi1999two}.

First we note that the MADM can be written in terms of nearest neighbor Hamiltonian densities as
\begin{equation}
  \mathcal{H}^{\text{SW}}=\sum_{k\in\Z} \mathcal{H}^{\text{SW}}_{k,k+1}
\end{equation} 
The components of the Hamiltonian density for a given number of excitations $M$ then read
\begin{equation}\label{eq:xxzham}
\left(\mathcal{H}^{\text{SW}}_{k,k+1}\right)_{ij}=q\left(\sum_{\substack{l=1\\l\neq i}}^{M+1}\frac{q^{i-l}}{[|i-l|]_q}\delta_{i,j}-\frac{q^{j-i}}{[|i-j|]_q}\delta_{i\neq j}\right)
\end{equation} 
where we use the conventions $[x]_q=(q^x-q^{-x})/(q-q^{-1})$ and $i,j=1,2,\ldots,M+1$. Taking $q\to 1$ we immediately obtain
\begin{equation}
 \lim_{q\to 1}\left(\mathcal{H}^{\text{SW}}_{k,k+1}\right)_{ij}=\sum_{\substack{l=1\\l\neq i}}^M\frac{1}{|i-l|}\delta_{i,j}-\frac{1}{|i-j|}\delta_{i\neq j}
\end{equation} 
which can be identified with the Hamiltonian density in \eqref{eq:harmonicaction} after noting that 
\begin{equation}
 \sum_{\substack{k=1\\k\neq i}}^{M+1}\frac{1}{|i-k|}=h(M-i+1)+h(i-1)\,.
\end{equation} 
Interestingly, the Hamiltonian density in \eqref{eq:xxzham} can be identified with the trigonometric version of the non-compact spin $\frac{1}{2}$ Heisenberg chain \cite{frassek00}.

\subsection{Scaling limit leading to integral form}
\label{scaling}
The Hamiltonian introduced in the Section \ref{spin12}
admits an integral representation that somehow
naturally generalizes the stochastic evolution in a discrete
space to a random dynamics in a continuum space.
This is obtained by considering the non-negative real
variable $x_i \in [0,\infty)$ defined by the scaling limit 
$x_i = \lim_{M\to\infty} x_i^{(M)}$ where 
$x_i^{(M)} = \frac{m_i}{M}$ is the fraction of
particles at site $i$.
Furthermore to have a meaningful limit
of the Hamiltonian boundary terms one needs to scale the
parameters as $\beta_i^{(M)} = 1 - \frac{\lambda_i}{M}$
with $\lambda_i \in (0,\infty)$, $i=1,N$.
With this procedure (details of computation are reported in  Appendix \ref{appA})
one moves from the generator $-\mathcal{H}^t$, with $\cal{H}$ given in \eqref{eq:fullham}, 
to
the generator $\mathcal{L}$ defined by
\be
\label{ham:levy}
\mathcal{L} = \mathcal{L}_1 + \sum_{i=1}^{N-1} \mathcal{L}_{i,i+1} + \mathcal{L}_N,
\ee
with bulk term
\begin{eqnarray}
\label{levy:bulk}
\mathcal{L}_{i,i+1} f(x_i,x_{i+1})
&= & 
\int_{0}^{x_i} \frac{d \alpha}{\alpha}\Big\{ f(x_i - \alpha, x_{i+1} + \alpha) - f(x_i,x_{i+1})\Big\}
\nonumber\\
&+ &
\int_{0}^{x_{i+1}} \frac{d \alpha}{\alpha}\Big\{ f(x_i + \alpha, x_{i+1} - \alpha) - f(x_i,x_{i+1})\Big\},
\end{eqnarray}
and boundary terms
\begin{eqnarray}
\label{levy:boundary}
\mathcal{L}_i f(x_i) 
&= & 
\int_{0}^{x_i} \frac{d \alpha}{\alpha}\Big\{ f(x_i - \alpha) - f(x_i)\Big\}
\nonumber \\
&+ &
\int_{0}^{\infty} d \alpha \frac{e^{-\lambda_i\alpha}}{\alpha}\Big\{ f(x_i + \alpha) - f(x_i)\Big\},
\end{eqnarray}
where $i=1,N$ and $0<\lambda_1,\lambda_N<\infty$. 
As the Markov generator $-\mathcal{H}^t$ is associated to an interacting particle
systems, the operator in  \eqref{ham:levy} is also the generator of a Markov process
$\{{\bf X}(t), \, t \ge 0\}$ that turns out to be a {\em L\'evy process}.
Here ${\bf X}(t) = (X_1(t),\ldots,X_N(t))$ and
the $i^{th}$ component $X_i(t)$ denotes the amount of 
a non-negative quantity (e.g. mass, energy, $\ldots$)
at site $i \in \{1,\ldots,N\}$ at time $t\ge 0$.
The process $\{{\bf X}(t), \, t \ge 0\}$ is a pure-jump process described as follows.
In the bulk, jumps of size $[\alpha, \alpha+d\alpha)$,
that move mass across the edge $(i,i+1)$ symmetrically, 
occur as Poisson process with intensity $\frac{d\alpha}{\alpha}$.
At the left boundary (site $1$),
negative jumps decreasing the mass 
by an amount in the range $[\alpha, \alpha+d\alpha)$
occur with intensity $\frac{d\alpha}{\alpha}$, and 
positive jumps increasing the mass by the same amount
occur with intensity $\frac{d\alpha}{\alpha} e^{-\lambda_1 \alpha}$
with $\lambda_1 \in (0,\infty)$.
Similar jumps occur at the right boundary (site $N$),
now with a parameter $\lambda_N \in (0,\infty)$.

One can show that in the particular case $\lambda_1 = \lambda_N = \lambda$
the product measure with marginal the Exponential distribution, i.e. 
the measure with density
\be
p(x_1, \ldots, x_N) = \prod_{i=1}^N \lambda e^{-\lambda x_i} 1_{\{x_i >0\}}, 
\ee
is reversible. This can be proved by showing that
the generator $\mathcal{L}$ in  \eqref{ham:levy}
is self-adjoint in the Hilbert space $L_2(p(\cdot),dx)$.
The generic case $\lambda_1 \neq  \lambda_N$ yields 
a boundary driven  {\em non-equilibrium} L\'evy process.

The bulk generator in \eqref{levy:bulk} reminds of the 
bulk Hamiltonian considered by  Derkachov in \cite{Derkachov:1999pz}
which reads
\begin{eqnarray}
\label{beis}
{H}_{i,i+1} f(x_i,x_{i+1})
&= & 
\int_{0}^{1} \frac{d \alpha}{\alpha}\Big\{ f((1-\alpha)x_i + \alpha x_{i+1}, x_{i+1}) - f(x_i,x_{i+1})\Big\}
\nonumber\\
&+ &
\int_{0}^{1} \frac{d \alpha}{\alpha}\Big\{ f(x_i,\alpha x_i + (1-\alpha) x_{i+1}) - f(x_i,x_{i+1})\Big\}.
\end{eqnarray}
This indeed corresponds to another representation of the Hamiltonian \eqref{ham:fadeev}.
The relation between \eqref{eq:harmonicaction} and \eqref{beis} is obtained by acting on polynomials
\be
f(x_i,x_{i+1}) = x_i^{m_i} x_{i+1}^{m_{i+1}} \qquad
\text{and}  \qquad
|m_i,m_{i+1}\rangle =x_i^{m_i} x_{i+1}^{m_{i+1}},
\ee
one has
\be
{H}_{i,i+1} f(x_i,x_{i+1}) =- \mathcal{H}_{i,i+1} |m_i,m_{i+1}\rangle .
\ee
Further, the relation between \eqref{levy:bulk} and \eqref{beis} is that,
defining 
\be
f_1(x_i,x_{i+1}) = x_i^{m_i} x_{i+1}^{m_{i+1}} \qquad
\text{and}  \qquad
f_2(x_i,x_{i+1}) =\frac{x_i^{m_i}}{(m_i)!}\frac{ x_{i+1}^{m_{i+1}}}{(m_{i+1})!},
\ee
one has
\be
{H}_{i,i+1} f_1(x_i,x_{i+1}) =\left[ \mathcal{L}_{i,i+1} f_2(\tilde x_i,\tilde x_{i+1})\right]_{\tilde x^{m}\to m! x^{m}} .
\ee
after the  subsequent replacement $\tilde x^{m}\to m! x^{m}$ on the right hand side.

\subsection{General spin and relation to q-Hahn zero range process}
\label{povo}
The Hamiltonian for the general spin $s$ non-compact Heisenberg chain can be obtained from the integral formula given in \cite{Derkachov:1999pz} by acting on polynomials, see also \cite{Martins:2009dt}. 
The local action reads
\begin{equation}\label{eq:hacts}
\begin{split}
 \mathcal{H}_{i,i+1}^{(s)}|m_i\rangle\otimes|m_{i+1}\rangle&=\left(h^{(s)}(m_i)+h^{(s)}(m_{i+1})\right)|m_i\rangle\otimes|m_{i+1}\rangle\\
 &\qquad-\sum_{k=1}^{m_i}\frac{1}{k}\frac{\Gamma (m_i+1) \Gamma (m_i-k+2 s)}{ \Gamma (m_i-k+1) \Gamma (m_i+2 s)}|m_i-k\rangle\otimes|m_{i+1}+k\rangle
 \\&\qquad-\sum_{k=1}^{m_{i+1}}\frac{1}{k}\frac{\Gamma (m_{i+1}+1) \Gamma (m_{i+1}-k+2 s)}{\Gamma (m_{i+1}-k+1)\Gamma (m_{i+1}+2 s)  }|m_i+k\rangle\otimes|m_{i+1}-k\rangle
 \end{split}
\end{equation} 
with
\begin{equation}
 h^{(s)}(m)=\sum_{k=1}^m\frac{1}{k+2s-1}\,.
\end{equation} 
Here the action of the $\mathfrak{sl}(2)$ generators \eqref{eq:sl2com} on each site of the spin chain is given by 
\begin{equation}\label{eq:sl2spinsaction}
S_+|m\rangle=(m+2s)|m+1\rangle\,,\quad S_-|m\rangle =m|m-1\rangle\,,\quad S_0|m\rangle =(m+s)|m\rangle 
\end{equation} 
The Hamiltonian density in  \eqref{eq:hacts} can be written in the standard form as
\begin{equation}
\label{ham:fadeevs}
 \mathcal{H}_{i,i+1}=2(\psi(\KK_{i,i+1})-\psi(2s))\,,
\end{equation} 
which is a generalization of \eqref{ham:fadeev}. The action of $\KK_{i,i+1}$ is determined via the tensor product decomposition for general spin $s$
\begin{equation}\label{eq:irreps}
 [s]\otimes [s]=\bigoplus_{j=0}^\infty[\Lambda_j]
\end{equation} 
where $[\kk]$ denotes the representation with representation label $\kk$ and $\Lambda_j=2s+j$.
If we set $s=\frac{1}{2}$ we recover \eqref{eq:irreps12}.

The Hamiltonian density in  \eqref{eq:hacts} is stochastic. To our knowledge, the process with Hamiltonian density \eqref{eq:hacts} has not been considered so far, not even on the closed chain.
It is not difficult to check that for the closed chain the particle process $\{{\bf M}^{(s)}(t)\, , \, t\ge 0\}$ with Hamiltonian
\begin{equation}
\label{Hs}
  \mathcal{H}^{(s)}=\sum_{i=1}^{N} \mathcal{H}^{(s)}_{i,i+1}
  \quad\text{with}\quad  \mathcal{H}^{(s)}_{N,N+1}= \mathcal{H}^{(s)}_{N,1}
\end{equation}  
has a invariant measure given by a product of Negative Binomial distribution 
with parameters $0<\beta<1$ and $2s >0 $. Namely, the law
\be
\mathbb{P}({\bf M}^{(s)}(t) = {\bf m}) = \prod_{i=1}^N \frac{\beta^{m_i}}{m_i!} \frac{\Gamma(m_i + 2s)}{\Gamma(2s)} (1-\beta)^{2s}
\ee
is reversible (and thus stationary). For $s=1/2$ we recover \eqref{geo}.
The generalisation of the integrable boundary terms \eqref{eq:harmbnd} reads
\begin{equation}\label{eq:harmbnds}
  \mathcal{H}_i|m_i\rangle=\left(h^{(s)}(m_i)+\sum_{k=1}^\infty\frac{\beta_i^k}{k}\right)|m_i\rangle -\sum_{k=1}^{m_i}\frac{1}{k}\frac{\Gamma (m_i+1) \Gamma (m_i-k+2 s)}{ \Gamma (m_i-k+1) \Gamma (m_i+2 s)}|m_i-k\rangle -\sum_{k=1}^\infty \frac{\beta_i^k}{k}|m_i+k\rangle ,
\end{equation} 
where $i=1,N$ and $0<\beta_1,\beta_N<1$.

Remarkably, the non-compact spin $s$ Hamiltonian density in \eqref{eq:hacts} is related to to the class of zero range models studied by Povolotsky in \cite{Povolotsky}
in the totally asymmetric context, and further extended by Barraquand and Corwin in \cite{barraquand} to the partially asymmetric case. More precisely we find that the transition rates 
\begin{equation}\label{eq:povrate}
 \varphi_{\mu,\nu,\gamma}(m|n)=\mu^m\frac{(\nu/\mu;\gamma)_m(\mu;\gamma)_{n-m}}{(\nu;\gamma)_n}\frac{(\gamma;\gamma)_n}{(\gamma;\gamma)_{n-m}(\gamma;\gamma)_m}\,,
\end{equation} 
where $m=0,1,\ldots,n$ and $(a;\gamma)_m=\prod_{j=0}^{m-1}(1-a\gamma^j)$, reduce to the ones in \eqref{eq:hacts} in the limit
\begin{equation}
 \lim_{s'\to s}  \frac{1}{2(s-s')}\lim_{\gamma\to 1} \varphi_{\gamma^{2s},\gamma^{2s'},\gamma}(m|n)=-\frac{1}{m}\frac{\Gamma (n+1) \Gamma (n-m+2 s)}{ \Gamma (n-m+1) \Gamma (n+2 s)}\,.
\end{equation}

The rational case studied here and the results for the trigonometric case in  \cite{frassek00} suggest that the transfer matrix constructed from the stochastic R-matrix for $U_q(A_1^{(1)})$ for the continuous time Markov chain  in \cite{Kuniba} is related to Baxter's Q-operator of the non-compact spin-$s$ XXZ chain in the closed setting. The transfer matrix in \cite{Kuniba} has two special points which yield a left and a right moving TAZRP. These can then be combined into a process of the type which we discussed here with particles hopping to the left and to the right.
A similar relation between the local charges of the Q-operator, which also arise from two special points, and the spin chain Hamiltonian was obtained in \cite[Appendix C]{Frassek:2012mg} for the rational limit with $s=\frac{1}{2}$ based on the oscillator construction of Q-operators \cite{Bazhanov:2010ts,Bazhanov:2010jq,Frassek:2011aa}, see also \cite{Frassek:2010ga,Frassek:2017bfz} for the Q-operator construction of supersymmetric spin chains including the one that underlies $\mathcal{N}=4$ SYM at weak coupling.

\section{Quantum inverse scattering method}
\label{sec:qism}
In this section we describe the spin chain defined by the nearest neighbor Hamiltonian \eqref{eq:fullham} within the quantum inverse scattering method for boundary integrable models \cite{Sklyanin:1988yz}. We construct the fundamental transfer matrix $T(x)$ with the infinite-dimensional spin $s=\frac{1}{2}$ representation in the auxiliary space and the transfer matrix with the two-dimensional fundamental representation in the auxiliary space $T_\square(x)$ for the most general, non-diagonal, boundary conditions. The corresponding K-matrices for the case of the fundamental transfer matrix appear to be new. They are expressed in terms of generators of $\mathfrak{sl}(2)$. The two different transfer matrices commute with each other and belong to the commuting family of operators of the integrable spin chain. Both of them play a distinct role in the quantum inverse scattering method. While the fundamental transfer matrix yields the Hamiltonian, the transfer matrix with two-dimensional auxiliary space is commonly used in the framework of the algebraic Bethe ansatz in order to diagonalise the family of commuting operators, see \cite{Faddeev:1996iy} for an excellent review. After specifying to appropriate boundary conditions we extract the Hamiltonian \eqref{eq:fullham} from the fundamental transfer matrix and thus find that it is integrable. This was known for the closed chain, see in particular  \cite{Beisert:2003jj}. The open chain has  been studied for the case of diagonal (identity) boundary conditions in \cite{Derkachov:1999ze,Derkachov:2003qb} and for a triangular case in  \cite{Belitsky:2014rba}.
Finally, we discuss how the Hamiltonian \eqref{eq:fullham} can be diagonalised using the algebraic Bethe ansatz.
\subsection{Construction of the transfer matrix with two-dimensional auxiliary space}
In order to construct the fundamental transfer matrix which contains the information about the Hamiltonian we first define the transfer matrix with the two-dimensional representation in the auxiliary space. It can be defined as the trace
\begin{equation}\label{eq:batrans}
 T_\square(x)=\tr_\square K(x)U(x)\,,
\end{equation} 
where $U(x)$ denotes the double-row monodromy
\begin{equation}
 U(x)=L_1(x)L_2(x)\cdots L_N(x)\hat K(x)L_N(x)\cdots L_2(x)L_1(x)\,.
\end{equation} 
Here we introduced the spectral parameter $x\in \mathbb{C}$ and the Lax matrix 
\begin{equation}\label{eq:lax}
 L_i(x)=\left(\begin{array}{cc}
                         x+\frac{1}{2}+S_0^{[i]}&-S_-^{[i]}\\
                         S_+^{[i]}&x+\frac{1}{2}-S_0^{[i]}
                        \end{array}
\right)
\end{equation} 
with the generators $S_\pm^{[i]}$ and $S_0^{[i]}$ acting on spin chain site $i$ and the most general $2\times 2$ K-matrices
\begin{equation}\label{eq:boldK}
  K(x) =\left(\begin{array}{cc}
                         p_1+p_2(x+1)&p_3(x+1)\\
                         p_4(x+1)&p_1-p_2(x+1)
                        \end{array}
\right)\,,\qquad 
  \hat K(x) =\left(\begin{array}{cc}
                         q_1+xq_2&xq_3\\
                         xq_4&q_1-xq_2
                        \end{array}
\right)\,,
\end{equation} 
see \cite{deVega:1992zd}. Both of the K-matrices depend on four complex variables $p_i$ and $q_i$ respectively with $i=1,\ldots,4$. We note that one degree of freedom can be absorbed into an overall normalisation. It will however be convenient for us to keep it.
The transfer matrix defined in this way commutes at different values of the spectral parameter
\begin{equation}\label{eq:comt}
 [ T_\square(x), T_\square(y)]=0\,.
\end{equation} 
This can be verified using the standard R-matrix $ R(z)=z+P$
where $P=\sum_{a,b=1}^2e_{ab}\otimes e_{ba}$, with $(e_{ab})_{cd}=\delta_{ac}\delta_{bd}$, denotes the permutation matrix. 
The commutativity \eqref{eq:comt} then follows from the Yang-Baxter equation
\begin{equation}
 R(x-y)\left(L(x)\otimes \ID\right)\left(\ID\otimes L(y)\right)=\left(\ID\otimes L(y)\right)\left(L(x)\otimes \ID\right)R(x-y)\,,
\end{equation} 
the boundary Yang-Baxter equation 
\begin{equation}
R(x-y)\left(\hat K(x)\otimes\ID\right)R(x+y)\left(\ID\otimes\hat K(y)\right)=\left(\ID\otimes\hat K(y)\right)R(x+y)\left(\hat K(x)\otimes\ID\right)R(x-y)\,,
\end{equation} 
and its corresponding version for $K$. Here $\ID$ denotes the $2\times2$ identity matrix. See \cite{Sklyanin:1988yz} for further details. 

In the next subsection we construct the fundamental transfer matrix which contains the physical information about the spin chain and commutes with the transfer matrix $T_\square$.

\subsection{Fundamental transfer matrix}
For the fundamental transfer matrix the representation of the auxiliary space coincides with the one at a single site of the quantum space, cf.~\eqref{eq:sl2action}. It can be defined in analogy to $T_\square$ via 
\begin{equation}\label{eq:fundtrans}
 T(z)=\tr_{a}\mathcal{K}(z)\mathcal{U}(z)\,.
\end{equation} 
Here the trace is taken in the infinite-dimensional auxiliary space $a$ and the double-row monodromy 
\begin{equation}\label{eq:dmon}
 \mathcal{U}(z)=\mathcal{R}_1(z)\mathcal{R}_2(z)\cdots \mathcal{R}_N(z)\hat{\mathcal{K}}(z)\mathcal{R}_N(z)\cdots \mathcal{R}_2(z)\mathcal{R}_1(z)\,.
\end{equation} 
Here $\mathcal{R}_i$ acts non-trivially in space $i$ of the quantum space and the auxiliary space $a$ which is suppressed in our notation.
We stress again, that the difference between $T_\square$ and $T$ is that $T_\square$ is constructed as the trace of a two-dimensional auxiliary space while one of $T$ is infinite-dimensional.

The R-matrix $\mathcal{R}(z)$ in \eqref{eq:dmon} is well known, see \cite{Kulish:1981gi} as well as \cite{Faddeev:1996iy}. It can be written in terms of $\Gamma$-functions as 
\begin{equation}\label{eq:rmat}
 \mathcal{R}(x)=(-1)^\mathbb{S}\frac{\Gamma(2s-x)}{\Gamma(2s+x)}\frac{\Gamma(\mathbb{S}+x)}{\Gamma(\mathbb{S}-x)}\,.
\end{equation} 
where the operator $\KK$ is the same that appeared in the Hamiltonian \eqref{ham:fadeev}.
Thus the eigenvalues of the R-matrix \eqref{eq:rmat} can easily be obtained for any irreducible representation of the tensor product decomposition \eqref{eq:irreps}.
Here the normalisation in \eqref{eq:rmat} ensures that $\mathcal{R}(0)=(-1)^\mathbb{S}$ which can be interpreted as the permutation operator.

The R-matrix \eqref{eq:rmat} satisfied the Yang-Baxter equation
\begin{equation}
 \mathcal{R}(x-y)\left(L(x)\otimes \ID\right)\left(\ID\otimes L(y)\right)=\left(\ID\otimes L(y)\right)\left(L(x)\otimes \ID\right)\mathcal{R}(x-y)\,.
\end{equation} 
Further, demanding that the fundamental transfer matrix \eqref{eq:fundtrans} commutes with the transfer matrix \eqref{eq:batrans}, i.e.
\begin{equation}\label{eq:commutet}
 [T(x),T_\square(y)]=0
\end{equation} 
we obtain  the boundary Yang-Baxter equations for $\mathcal{K}$ and $\hat{\mathcal{K}}$ for the left and right boundary respectively. They involve the K-matrices \eqref{eq:boldK} and read
\begin{equation}\label{eq:bybe1}
 L(x-y)\hat{\mathcal{K}}(x)L(x+y)\hat K(y)=\hat K(y)L(x+y)\hat{\mathcal{K}}(x)L(x-y)
\end{equation} 
and 
\begin{equation}\label{eq:bybe2}
 L(y-x)\mathcal{K}(x)L(-x-y-2)K(y)=K(y)L(-x-y-2)\mathcal{K}(x)L(y-x)
\end{equation} 
Here we note that the Lax matrix \eqref{eq:lax} satisfies the unitarity relations
\begin{equation}\label{eq:unit}
 L(x)L(-x)=\left(x+s-\frac{1}{2}\right)\left(-x+s-\frac{1}{2}\right)\,,\quad  L^t(x)L^t(-x-2)=\left(x+s+\frac{1}{2}\right)\left(-x+s-\frac{3}{2}\right)\,,
\end{equation} 
where $L^t$ denotes the Lax matrix transposed in the $2\times 2$ auxiliary space.

An expression in terms of $\mathfrak{sl}(2)$ generators is unknown for the off-diagonal case, see \cite{Baseilhac:2017hoz} for the diagonal case which has been studied in the context of Q-operators \cite{Frassek:2015mra,Baseilhac:2017hoz}. In the next subsection, we will derive the solutions  $\mathcal{K}$ and $\hat{\mathcal{K}}$ of the boundary Yang-Baxter equations \eqref{eq:bybe1} and \eqref{eq:bybe2}.

\subsection{General solution to the boundary Yang-Baxter equation}
In the following we first obtain the solution $\hat{\mathcal{K}}$ to the boundary Yang-Baxter equation \eqref{eq:bybe1}. As we will see the solution ${\mathcal{K}}$ then follows from $\hat{\mathcal{K}}$.

First we note that the boundary Yang-Baxter equation \eqref{eq:bybe1} is equivalent to the conditions
\begin{equation}\label{eq:level1}
 \left[\left\{L(x),Q\right\},\hat{\mathcal{K}}(x)\right]=0\,,
\end{equation} 
and 
\begin{equation}\label{eq:level2}
 2q_1[L(x),\hat{\mathcal{K}}(x)]=\left[Q,L(x)\hat{\mathcal{K}}(x)L(x)\right]\,.
\end{equation} 
Here we defined the matrix 
\begin{equation}
 Q=\left(\begin{array}{cc}
               q_2&q_3\\
               q_4&-q_2
              \end{array}
\right)\,.
\end{equation} 
One finds that the first equation \eqref{eq:level1} is equivalent to the condition 
\begin{equation}\label{eq:level11}
 \left[2q_2 S_0+q_3S_+-q_4S_-,\hat{\mathcal{K}}(x)\right]=0
\end{equation} 
and the second \eqref{eq:level2} can be written as
\begin{equation}\label{eq:level22}
 \hat{\mathcal{K}}(x)\left(2q_1L(-x)+L(x)QL(-x)\right)=\left(2q_1L(-x)+L(-x)QL(x)\right)\hat{\mathcal{K}}(x)
\end{equation} 
using the  unitarity relation \eqref{eq:unit}. 
In order to find a solution to these equations we note that the operator in \eqref{eq:level11} can be diagonalised via
\begin{equation}
 e^{\alpha S_-} e^{-\beta S_+}\left[2q_2S_0+q_3S_+-q_4 S_-\right] e^{\beta S_+} e^{-\alpha S_-}=\gamma S_0
\end{equation} 
after introducing the  parametrisation 
\begin{equation}\label{eq:idbnd1}
 q_1=\de \,,\quad q_2=\frac{1}{2}(1+2\alpha\beta)\gamma\,,\quad q_3=-(1+\alpha\beta)\beta\gamma\,,\qquad q_4=\alpha \gamma\,.
\end{equation} 
This motivates the ansatz for the K-matrix
\begin{equation}\label{eq:offK}
\hat{\mathcal{K}}(x)= e^{\beta S_+}\, e^{-\alpha S_-}\,\hat{\mathcal{K}}_0(S_0;x)\, e^{\alpha S_-}\, e^{-\beta S_+}\,,
\end{equation} 
where the middle term $\hat{\mathcal{K}}_0(S_0;x)$ only depends on the generator $S_0$ and not on $S_\pm$.
Substituting this ansatz into \eqref{eq:level22} we obtain the difference equation
\begin{equation}
 \frac{\hat{\mathcal{K}}_0(S_0;x)}{\hat{\mathcal{K}}_0(S_0-1;x)}=\frac{(S_0-\frac{1}{2}+x)\gamma+2\de}{(S_0-\frac{1}{2}-x)\gamma+2\de}\,.
\end{equation} 
This equation is solved by the fraction of $\Gamma$-functions
\begin{equation}\label{eq:diagK}
\hat{\mathcal{K}}_0(S_0;x)=\frac{\Gamma(\frac{1}{2}+s+2\frac{\de}{\gamma}-x)}{\Gamma(\frac{1}{2}+s+2\frac{\de}{\gamma}+x)}\frac{\Gamma\left(\frac{1}{2}+S_0+2\frac{\de}{\gamma}+x\right)}{\Gamma\left(\frac{1}{2}+S_0+2\frac{\de}{\gamma}-x\right)}
\end{equation} 
up to a function periodic in $S_0$. The normalisation is chosen such that $\hat{\mathcal{K}}(0)=\ID$. To our knowledge the solution \eqref{eq:offK} with \eqref{eq:diagK} is new.

The K-matrix for the other boundary can be obtained from the relation 
\begin{equation}
 \mathcal{K}(x)=\frac{1}{\hat{\mathcal{K}}(x+1)}\,,
\end{equation} 
and renaming the parameters appearing.
We find that it can be written explicitly as
\begin{equation}
\mathcal{K}(x)=e^{\beta' S_+}\, e^{-\alpha' S_-}\,{\mathcal{K}}_0(S_0;x)\, e^{\alpha' S_-}\, e^{-\beta' S_+}
\end{equation} 
with
\begin{equation}\label{eq:diagK2}
{\mathcal{K}}_0(S_0;x)=\frac{\Gamma(\frac{3}{2}+s+2\frac{\de'}{\gamma'}+x)}{\Gamma(-\frac{1}{2}+s+2\frac{\de'}{\gamma'}-x)}\frac{\Gamma\left(-\frac{1}{2}+S_0+2\frac{\de'}{\gamma'}-x\right)}{\Gamma\left(\frac{3}{2}+S_0+2\frac{\de'}{\gamma'}+x\right)}
\end{equation} 
Here we used the same parametrisation as in \eqref{eq:idbnd1} for the variables $p_i$, i.e. 
\begin{equation}\label{eq:idbnd2}
 p_1=-\de'\,,\quad p_2=\frac{1}{2}(1+2\alpha'\beta')\gamma'\,,\quad p_3=-(1+\alpha'\beta')\beta'\gamma'\,,\qquad p_4=\alpha' \gamma'\,.
\end{equation} 
The extra minus sign in $p_1$ arises from the inversion relation of the K-matrices $K^{-1}(x)\sim K(-x)$ that is used  when identifying \eqref{eq:bybe1} and \eqref{eq:bybe2}.
Finally we note that the K-operators $\hat{\mathcal{K}}$ and $\mathcal{K}$ only depend on the ratio $\delta/\gamma$ and $\delta'/\gamma'$. In the following we will be interested in the case where $2\delta=s-\frac{1}{2}$, $2\delta'=\frac{1}{2}-s$  and $\gamma,\gamma'\neq 0$. The latter will appear as an overall factor in the K-matrices \eqref{eq:boldK}, c.f.~\eqref{eq:kfunds}, and is set to $\gamma=\gamma'=1$.

In the next subsection we derive the Hamiltonian \eqref{eq:fullham} from the fundamental transfer matrix \eqref{eq:fundtrans} for a special choice of boundary parameters \eqref{eq:idbnd1} and \eqref{eq:idbnd2}.

\subsection{Derivation of the stochastic boundary terms}\label{sec:devH}
In this section we compute the Hamiltonian for $s=\frac{1}{2}$ by taking the logarithmic derivative of the fundamental transfer matrix at $x=0$, see \cite{Sklyanin:1988yz}. The latter can be written in terms of the R-matrices and K-matrices in \eqref{eq:fundtrans} as
\begin{equation}\label{eq:logderiv}
\frac{\partial}{\partial x}\ln T(x)\big|_{x=0}=\frac{\tr_a \mathcal{K}_a'(0)}{\tr_a \mathcal{K}_a(0)}+2\frac{\tr_a \mathcal{K}_a(0)\mathcal{H}_{a,1}}{\tr_a \mathcal{K}_a(0)}+\frac{\hat{\mathcal{K}}_N'(0)}{\hat{\mathcal{K}}_N(0)}+2\sum_{k=1}^{N-1}\frac{\partial}{\partial x}\ln \mathcal{R}_{k,k+1}(x)\big|_{x=0}\,,
\end{equation} 
where the subscript $a$ denotes an infinite-dimensional auxilliary space and $\hat{\mathcal{K}}_N$ the K-matrix at the $N$-th site of the spin chain while ${\mathcal{K}}_a$ the K-matrix acting non-trivially in the auxiliary space.

In the following we will show that the logarithmic derivative at $x=0$ in \eqref{eq:logderiv} coincides with the Hamiltonian \eqref{eq:fullham} up to a constant, i.e. 
\begin{equation}\label{eq:hami}
\mathcal{H}=\frac{1}{2}\left(1+\frac{\partial}{\partial x}\ln T(x)\big|_{x=0}\right)\,,
\end{equation} 
when imposing the conditions
\begin{equation}\label{eq:identify}
 \alpha=\frac{1}{1-\beta}=\sum_{k=0}^\infty \beta^k\,,\qquad\alpha'=\frac{1}{1-\beta'}=\sum_{k=0}^\infty \beta'^k\,,\qquad \de=\de'=0\,,
\end{equation} 
where $0< \beta,\beta'<1$. Here, the constant term in \eqref{eq:hami} does not spoil the commutativity with the fundamental transfer matrix $T$ and $T_\square$. 
The logarithmic derivative of the R-matrix at the permutation point is straightforward to compute and yields the Hamiltonian density in \eqref{ham:fadeev}, i.e.
\begin{equation}
\mathcal{H}_{i,i+1}= \frac{\partial}{\partial x}\ln \mathcal{R}_{i,i+1}(x)\big|_{x=0}=2(\psi(\KK_{i,i+1})-\psi(1))
\end{equation} 
We stress  again that it was argued in \cite{Beisert:2003jj} that this expression is equivalent to the bulk action  \eqref{eq:harmonicaction}. 
The identification of the boundary terms $\mathcal{H}_1$ and $\mathcal{H}_N$ is given in the following two subsections. 
The boundary terms for general spin $s$ in \eqref{eq:harmbnds} can be derived analogously with $2\delta=s-\frac{1}{2}$, $2\delta'=\frac{1}{2}-s$ and $\gamma=\gamma'=1$.
\subsubsection{The right boundary}\label{sec:rbd}
First we note that that the logarithmic derivative of the K-matrix for $\delta=0$ at $x=0$  can be written as
\begin{equation}\label{eq:term1}
  \frac{\hat{\mathcal{K}}'(0)}{2\hat{\mathcal{K}}(0)}=e^{\beta S_+}\, e^{-\alpha S_-}\,\psi(S_0+\frac{1}{2})\, e^{\alpha S_-}\, e^{-\beta S_+}-\psi(1)\,.
\end{equation} 
To derive the action of this operator on a state $|m_N\rangle$ we evaluate the matrix elements
\begin{equation}
 \mathcal{O}=e^{\beta S_+}\, e^{-\alpha S_-}\,\psi(S_0+\frac{1}{2})\, e^{\alpha S_-}\, e^{-\beta S_+}\,.
\end{equation} 
This can be done by inserting two identity operators into $\mathcal{O}$ as follows. We find
\begin{equation}\label{eq:op}
\begin{split}
 \langle k|\mathcal{O}|l\rangle&=\sum_{m_1,m_2=0}^\infty  \langle k|e^{\beta S_+}\,|m_1\rangle\langle m_1| e^{-\alpha S_-}\psi\left(S_0+\frac{1}{2}\right)e^{\alpha S_-}\,|m_2\rangle \langle m_2|e^{-\beta S_+}|l\rangle\\
 &=\sum_{m_1,m_2=0}^\infty \binom{k}{m_1}\binom{m_2}{l}\beta^{k-m_1}(-\beta)^{m_2-l} \langle m_1| e^{-\alpha S_-}\psi\left(S_0+\frac{1}{2}\right)e^{\alpha S_-}\,|m_2\rangle\,,
 \end{split}
\end{equation} 
where  we have  used the relation 
\begin{equation}
\label{factorial}
 \langle m|e^{\gamma S_+}|l\rangle=\gamma^{m-l}\binom{m}{l}\,.
\end{equation} 
Next we note that the last part in \eqref{eq:op} yields
\begin{equation}\label{eq:midpart}
 \langle m_1| e^{-\alpha S_-}\psi\left(S_0+\frac{1}{2}\right)e^{\alpha S_-}\,|m_2\rangle=\psi(m_1+1)\delta_{m_1,m_2}-\sum_{r=1}^{m_2}\frac{\alpha^r}{r}\delta_{m_1,m_2-r}\,.
\end{equation} 
Now we can evaluate the matrix elements of $\mathcal{O}$ in \eqref{eq:op} by substituting \eqref{eq:midpart}. The diagonal term in \eqref{eq:midpart} yields
\begin{equation}\label{eq:coef1}
 \sum_{m=0}^\infty \binom{k}{m}\binom{m}{l}\beta^{k-m}(-\beta)^{m-l}\psi(m+1)=
\begin{cases}
0,\quad \text{for}\quad k<l\\
\psi(k+1),\quad \text{for}\quad k=l\\
-\frac{1}{k-l}\beta^{k-l} \quad \text{for}\quad k>l
\end{cases}\,.
\end{equation} 
To evaluate the part that emerges from the non diagonal term in  \eqref{eq:midpart} we identify 
\begin{equation}
 \alpha=\frac{1}{1-\beta}=\sum_{q=0}^\infty \beta^q
\end{equation} 
with $0<\beta<1$. Then we find 
\begin{equation}\label{eq:coef2}
-\sum_{m_1=0}^\infty \sum_{m_2=m_1+1}^\infty \binom{k}{m_1}\binom{m_2}{l}\beta^{k-m_1}(-\beta)^{m_2-l}\frac{\alpha^{m_2-m_1}}{m_2-m_1}=
\begin{cases}
-\frac{1}{l-k},\quad \text{for}\quad k<l\\
\sum_{m=1}^\infty\frac{\beta^m}{m},\quad \text{for}\quad k=l\\
0\quad \text{for}\quad k>l
\end{cases}\,.
\end{equation} 
Combining the results in \eqref{eq:coef1} and \eqref{eq:coef2} we obtain the matrix elements of $\mathcal{O}$ in \eqref{eq:op}. We finally find that 
for the identification in \eqref{eq:identify} the operator in \eqref{eq:term1} acting on a state $|m_N\rangle$ can be written as
\begin{equation}\label{eq:lasth}
 \frac{\hat{\mathcal{K}}_N'(0)}{2\hat{\mathcal{K}}_N(0)}|m_N\rangle=\left(h(m_N)+\sum_{k=1}^\infty\frac{\beta^k}{k}\right)|m_N\rangle -\sum_{k=1}^{m_N} \frac{1}{k}|m_N-k\rangle -\sum_{k=1}^\infty \frac{\beta^k}{k}|m_N+k\rangle \,.
\end{equation}
Further details can be found in Appendix~\ref{ap:comp1} and \ref{ap:comp2}.
\subsubsection{The left boundary}
The computation of the matrix elements for the left boundary turns out to be more simple than for the right boundary. First noting that 
\begin{equation}
 \tr_a\mathcal{K}_a(x)=\frac{1+x}{1+2x}
\end{equation} 
we find that
\begin{equation}\label{eq:tracesk}
 \tr_a \mathcal{K}_a(0)=1\,,\qquad  \tr_a \mathcal{K}'_a(0)=-1\,.
\end{equation} 
This yields in particular the first term  of the logarithmic derivative in \eqref{eq:logderiv}.
The non-trivial part is to compute the matrix elements of 
$\tr_a \mathcal{K}_a(0)\mathcal{H}_{a,1}$. 
First we evaluate the matrix elements of $\mathcal{K}_a(0)$. The middle part becomes a projector on the Fock vacuum. We find that
\begin{equation}
 \langle k|\mathcal{K}_a(0)|l\rangle=\beta'^k\sum_{m=l}^\infty\binom{m}{l}\alpha'^m(-\beta')^{m-l}=\beta'^{k}(1-\beta')\,.
\end{equation} 
using \eqref{eq:binomial}. Thus we have
\begin{equation}
 \tr_a \mathcal{K}_a(0)\mathcal{H}_{a,1}=\sum_{m,n=0}^\infty \langle m|\mathcal{K}_a(0)|n\rangle\langle n|\mathcal{H}_{a,1}|m\rangle=(1-\beta')\sum_{m,n=0}^\infty  \beta'^{m}\langle n|\mathcal{H}_{a,1}|m\rangle\,.
\end{equation}
where all bra's and ket's are in the auxiliary space $a$.
Then acting on a state $|m_1\rangle$  in the first space $1$ we get 
\begin{equation}\label{eq:firsth}
\begin{split}
  \tr_a \mathcal{K}_a(0)\mathcal{H}_{a,1}|m_1\rangle&=\left(\sum_{k=1}^\infty\frac{\beta'^k}{k}+h(m_1)\right)|m_1\rangle-\sum_{k=1}^{m_1}\frac{1}{k}|m_1-k\rangle-\sum_{k=1}^\infty\frac{\beta'^k}{k}|m_1+k\rangle\,.
  \end{split}
\end{equation} 

We have thus evaluated all terms in \eqref{eq:logderiv} that are needed to compute the Hamiltonian via \eqref{eq:hami}, cf. \eqref{eq:lasth}, \eqref{eq:tracesk} and \eqref{eq:firsth}. Finally we obtain the Hamiltonian action proposed in \eqref{eq:fullham} with the bulk part \eqref{eq:harmonicaction} and the boundary terms \eqref{eq:harmbnd} setting $\beta=\beta_N$ and $\beta'=\beta_1$. We conclude that the Hamiltonian is integrable. The corresponding transfer matrices $T$ and $T_\square$ are obtained from \eqref{eq:fundtrans} and \eqref{eq:batrans} after identifying the parameters via \eqref{eq:identify}. Similar boundary conditions were studied for the Hamiltonian of the $SL(2,\mathbb{R})$ spin chain with  discrete  series  representation in the context of  heavy-light operators in QCD \cite{Braun:2018fiz} and flux tubes in the $\mathcal{N}=4$ SYM theory \cite{Belitsky:2019ygi}.

\subsection{Symmetries and Bethe ans\"atze}

In this section we discuss the symmetries of the spin chain transfer matrix \eqref{eq:batrans} and how the algebraic Bethe ansatz can be applied. For simplicity we restrict ourselves to $s=\frac{1}{2}$. We find that for the equilibrium case the K-matrices can be diagonalised and that the general case it can be brought to a triangular form by a similarity transformation in the quantum space.

The transfer matrix $T_\square$, which is commonly used for the algebraic Bethe ansatz, can be obtained from the general expression \eqref{eq:batrans} under the identifications \eqref{eq:idbnd1}, \eqref{eq:idbnd2} and \eqref{eq:identify}. It commutes with the stochastic Hamiltonian \eqref{eq:fullham} and with the corresponding  fundamental transfer matrix introduced in Section~\ref{sec:devH}.
After the identification of the parameters, the K-matrices \eqref{eq:boldK} in the transfer matrix $T_\square(x)$ are of the form 
\begin{equation}\label{eq:kfunds}
\hat K(x)=\frac{x\gamma}{1-\beta}
 \left(
\begin{array}{cc}
 \frac{(\beta +1) }{2} & -\beta \\
 1& -\frac{(\beta +1)  }{2} \\
\end{array}
\right)\,,\qquad K(x)=\frac{(x+1)\gamma'}{1-\beta'}
 \left(
\begin{array}{cc}
 \frac{(\beta' +1) }{2 } & -\beta' \\
 1 & -\frac{(\beta' +1)  }{2 } \\
\end{array}
\right)\,,
\end{equation} 
with $\beta=\beta_1$ and $\beta'=\beta_N$. The bulk part remains unchanged.

\subsubsection{Equilibrium}
In the case of equilibrium, i.e. $\beta=\beta'$, we find that the K-matrices in the corresponding transfer matrices can be diagonalised by a similarity transformation which can be absorbed in the quantum space. 

We note that the K-matrices in \eqref{eq:kfunds} obey the relations
\begin{equation}
\hat K'(x)=S_\beta\hat K(x)S^{-1}_\beta=\frac{x\gamma  }{2}\sigma_3\,,\qquad  K'(x)=S_{\beta'} K(x)S^{-1}_{\beta'}=\frac{(z+1)\gamma}{2}\sigma_3\,,
\end{equation} 
with the Pauli matrix $\sigma_3=\diag(+1,-1)$. The similarity transformation reads 
\begin{equation}
 S_\beta=\left(
\begin{array}{cc}
1& -\beta \\
 \frac{1}{\beta-1}& \frac{1}{1-\beta} \\
\end{array}
\right)\,.
\end{equation} 
We further find that the Lax matrix enjoys the property
\begin{equation}
\begin{split}
S_\beta
 L(x)S_\beta^{-1}&=e^{\beta S_+}e^{\frac{1}{\beta-1}S_-}L(x)e^{-\frac{1}{\beta-1}S_-}e^{-\beta S_+}\,,
\end{split}
\end{equation} 
where we used the relations
\begin{equation}\label{eq:strans}
e^{\omega S_\pm}S_0e^{-\omega S_\pm}=S_0\mp \omega S_\pm\,,\qquad e^{\omega S_\pm}S_\mp e^{-\omega S_\pm}=S_\mp\mp2\omega S_0+\omega^2 S_\pm   \,.                                                                                                                                      \end{equation} 
As a consequence we can rewrite the transfer matrix \eqref{eq:fundtrans} for $\beta=\beta'$ as
\begin{equation}
 T_\square(z)=\mathcal{S}_\beta T'_\square(z)\mathcal{S}^{-1}_\beta \,,\qquad\text{with}\qquad \mathcal{S}_\beta=e^{\beta S_+}e^{\frac{1}{\beta-1}S_-}\otimes \ldots\otimes e^{\beta S_+}e^{\frac{1}{\beta-1}S_-}\,.
\end{equation} 
The transfer matrix $T'$ has diagonal boundary conditions and we expect that the  standard algebraic Bethe ansatz \cite{Sklyanin:1988yz} can be applied taking $|0\rangle\otimes\ldots\otimes|0\rangle$ as a reference state.

 \subsubsection{Non-equilibrium}\label{sec:baegen}
For general parameters $\beta$ and $\beta'$, the transfer matrix $T_\square$ in \eqref{eq:batrans} the K-matrices \eqref{eq:kfunds} can be brought to a triangular form. 

This can be seen as follows. 
First we note that the K-matrices \eqref{eq:kfunds} satisfy
\begin{equation}
 \hat K''(x)=
\left(
 \begin{array}{cc}
  1&-1\\
  0&1
 \end{array}\right)
 \hat K(x)
\left(
 \begin{array}{cc}
  1&1\\
  0&1
 \end{array}\right)=-x\gamma\left(
 \begin{array}{cc}
  \frac{1}{2}&0\\
  \frac{1}{\beta-1}& -\frac{1}{2}
 \end{array}\right)\,,
\end{equation} 
and 
\begin{equation}
  K''(x)=
\left(
 \begin{array}{cc}
  1&-1\\
  0&1
 \end{array}\right)
 K(x)
\left(
 \begin{array}{cc}
  1&1\\
  0&1
 \end{array}\right)=-(x+1)\gamma'\left(
 \begin{array}{cc}
  \frac{1}{2}&0\\
  \frac{1}{\beta'-1}& -\frac{1}{2}
 \end{array}\right)\,.
\end{equation} 
The change of the Lax matrices under this transformation can be absorbed into the quantum space. We have
\begin{equation}
\begin{split}
\left(
 \begin{array}{cc}
  1&-1\\
  0&1
 \end{array}\right)
 L(x)
\left(
 \begin{array}{cc}
  1&1\\
  0&1
 \end{array}\right)
 &=e^{S_+}L(x)e^{-S_+}\,,
\end{split}
\end{equation} 
using \eqref{eq:strans}. As a consequence we can write the transfer matrix in \eqref{eq:batrans} as
\begin{equation}
 T_\square(x)=S T''_\square(x)S^{-1}\,,\qquad\text{with}\qquad S=e^{S_+}\otimes \ldots\otimes e^{S_+}\,.
\end{equation} 

 Alternatively we can bring the K-matrices to an upper triangular form via
\begin{equation}\label{eq:u1}
 \hat K'''(x)=
\left(
 \begin{array}{cc}
  1&0\\
  -1&1
 \end{array}\right)
 \hat K(x)
\left(
 \begin{array}{cc}
  1&0\\
  1&1
 \end{array}\right)=x\gamma\left(
 \begin{array}{cc}
  \frac{1}{2}&-\frac{\beta}{1-\beta}\\
  0& -\frac{1}{2}
 \end{array}\right)\,,
\end{equation} 
and 
\begin{equation}\label{eq:u2}
  K'''(x)=
\left(
 \begin{array}{cc}
  1&0\\
  -1&1
 \end{array}\right)
 K(x)
\left(
 \begin{array}{cc}
  1&0\\
  1&1
 \end{array}\right)=(x+1)\gamma'\left(
 \begin{array}{cc}
  \frac{1}{2}&-\frac{\beta'}{1-\beta'}\\
  0& -\frac{1}{2}
 \end{array}\right)\,.
\end{equation} 
The change of the Lax matrices under this transformation can be absorbed into the quantum space. We have
\begin{equation}
\begin{split}
\left(
 \begin{array}{cc}
  1&0\\
  -1&1
 \end{array}\right)
 L(x)
\left(
 \begin{array}{cc}
  1&0\\
  1&1
 \end{array}\right)
 &=e^{-S_-}L(x)e^{S_-}\,,
\end{split}
\end{equation} 
using \eqref{eq:strans}. As a consequence we can write the transfer matrix in \eqref{eq:batrans} as
\begin{equation}
 T_\square(x)=S T'''_\square(x)S^{-1}\,,\qquad\text{with}\qquad S=e^{-S_-}\otimes \ldots\otimes e^{-S_-}\,.
\end{equation} 
Finally, we remark that as the K-matrices can be brought to a triangular form we expect the Baxter equation to be of the standard form, i.e. to coincide with the one of the spin chain with diagonal boundary conditions. 
The algebraic Bethe ansatz for these types of transfer matrices with triangular K-matrices and for finite-dimensional representations in the quantum space has been studied in \cite{Belliard2013,Antonio:2014qxa}. 
We plan to elaborate on the non-compact case, which is relevant here, elsewhere.

\section{Duality}
\label{sec:duality}

In this Section we show that the continuos-time Markov process defined by the Hamiltonian
\eqref{eq:fullham} has a {\em dual process} that allows to express the correlation
functions in terms of finitely many particles. To introduce such a process
we need to consider the enlarged quantum space $\widetilde{V}$  
which is defined as the $N+2$ 
fold tensor product 
\begin{equation}
\widetilde{V} =|\ell_0 \rangle \otimes|\ell_1\rangle \otimes\ldots \otimes|\ell_N\rangle \otimes|\ell_{N+1}\rangle.
\end{equation}  
Then the dual Hamiltonian is defined by 
\begin{equation}
\label{eq:fullhamdual}
{\widetilde{H}} = 
{\widetilde{\mathcal{H}}_{0,1}}
+\sum_{i=1}^{N-1}\mathcal{H}_{i,i+1} 
+ {\widetilde{\mathcal{H}}_{N,N+1}},
\end{equation} 
where the bulk part is given by \eqref{eq:fullham},
while the boundary terms reads
\begin{equation}\label{eq:harmbnddual}
  \widetilde{\mathcal{H}}_{0,1} |\ell_0\rangle \otimes |\ell_1\rangle = h(\ell_1) | \ell_0\rangle \otimes |\ell_1\rangle  -\sum_{k=1}^{\ell_1} \frac{1}{k} |\ell_0+k\rangle \otimes |\ell_1-k\rangle , 
\end{equation} 
and 
\begin{equation}\label{eq:harmbnddual2}
  \widetilde{\mathcal{H}}_{N,N+1} |\ell_N\rangle \otimes |\ell_{N+1}\rangle = h(\ell_N) | \ell_N\rangle \otimes |\ell_{N+1}\rangle  - \sum_{k=1}^{\ell_N} \frac{1}{k} |\ell_N-k\rangle \otimes |\ell_{N+1}+k\rangle.  
\end{equation} 
Thus, in the dual process, particles moves as in the original system while they are in the bulk; when
they reach the boundary sites they can be absorbed in {\em two extra sites}, called ${0}$ and ${N+1}$,
where they remain forever. As a consequence, in the long-time limit the dual process voids the chain: eventually 
all particles are absorbed either at the extra left site or at the extra right  site.
In Section \ref{dual-statement} we give a precise formulation of duality,
whose proof is then found in Section \ref{dual-proof}. 
Some consequences of duality are proved in Section \ref{dual-consequences}. 

\subsection{Duality function}
\label{dual-statement}
For a configuration  ${\boldsymbol m} = (m_1\ldots,m_N) \in {\mathbb{N}^{N}}$
and a dual configuration ${\boldsymbol\ell} = (\ell_0,\ell_1\ldots,\ell_N,\ell_{N+1}) \in {\mathbb{N}}^{N+2}$,
we define the duality function
\be
\label{duality:function}
D({\boldsymbol m} , {\boldsymbol\ell}) = \rho_a^{\ell_0} \cdot \Big[\prod_{i=1}^N   {m_i \choose \ell_i} \Big ] \cdot \rho_b^{\ell_{N+1}}
\ee
where
$$
\rho_a = \frac{\beta_1}{1-\beta_1},  \qquad\qquad  \rho_b = \frac{\beta_N}{1-\beta_N}.
$$
We denote by $\{{\bf M}(t), \, t \ge 0\}$ the original process defined by the Hamiltonian \eqref{eq:fullham} 
and by  $\{{\bf L}(t), \, t \ge 0\}$ the dual process with Hamiltonian \eqref{eq:fullhamdual}.
Here ${\bf M}(t) = (M_1(t),\ldots,M_N(t))$ and  ${\bf L}(t) = (L_0(t),L_1(t),\ldots,L_N(t), L_{N+1}(t))$.
Note that the dual process has two additional components. 
Duality is then expressed by the following equivalence of expectation values
\be
\label{dual:relation}
\mathbb{E}_{{\bf m}} \Big[D({\bf M}(t), {\boldsymbol \ell}) \Big]= \widetilde{\mathbb{E}}_{\boldsymbol \ell} \Big[D({\bf m}, {\bf L}(t))\Big],
\ee
where $\mathbb{E}_{\bf m}$ denotes expectation w.r.t. the original process started from
the configuration ${\bf m}$, and  $\widetilde{\mathbb{E}}_{\boldsymbol \ell}$ denotes expectation w.r.t. the 
dual process started from the configuration ${\boldsymbol \ell}$.
More explicitly, duality amounts to
\be
\sum_{{\bf m}'\in \mathbb{N}^N}  D({\bf m}', {\boldsymbol \ell}) \, \mathbb{P} ({\bf M}(t) = {\bf m}' \; | \; {\bf M}(0) = {\bf m}  ) = 
\sum_{{\boldsymbol \ell}' \in \mathbb{N}^{N+2}}  D({\bf m}, {\boldsymbol \ell}') \, \mathbb{P} ({\bf L}(t) = {\boldsymbol \ell'} \; | \;  {\bf L}(0) = {\boldsymbol \ell}).
\ee

\subsection{Proof of duality}
\label{dual-proof}
In bra-ket notation, the duality relation \eqref{dual:relation} reads
\be
\label{dual:braket}
\sum_{{\bf m}'\in \N^{N}} \langle {\bf m}' | D | {\boldsymbol \ell} \rangle \langle {\bf m}' | e^{-t \mathcal{H}} | {\bf m} \rangle =  \sum_{{\boldsymbol \ell}'\in \N^{N+2}} \langle {\bf m} | D | {\boldsymbol \ell}' \rangle \langle {\boldsymbol \ell}' | e^{-t\widetilde{\mathcal{H}}} | {\boldsymbol \ell} \rangle,
\ee
where $D$ is the matrix whose elements are the duality function in \eqref{duality:function}
\be
\langle {\bf m}' | D | {\boldsymbol \ell} \rangle = D({\bf m}',{\boldsymbol \ell})
\ee
and $e^{-t\mathcal{H}}$ is the semigroup of the original process whose elements $\langle {\bf m}' | e^{-t \mathcal{H}} | {\bf m} \rangle$ give 
the probability that the original process $\{{\bf M}(t), \,  t\ge 0\}$ is at configuration ${\bf m}'$
at time $t$,  having started from the configuration ${\bf m}$ at time $0$.
Similarly $e^{-t\widetilde{\mathcal{H}}}$ is the semigroup of the dual process $\{{\bf L}(t), \, t \ge 0\}$ .

Using in the left hand side of \eqref{dual:braket} the equality
\be
\langle {\bf m}' | e^{-t \mathcal{H}} | {\bf m} \rangle = \langle {\bf m} | e^{-t \mathcal{H}^t} | {\bf m}' \rangle,
\ee 
where $\mathcal{H}^t$ denotes the transpose of $\mathcal{H}$, and using the resolutions 
of the identity
\be
1 = \sum_{{\bf m}'\in \N^{N}}  | {\bf m}' \rangle \langle {\bf m}' | =  \sum_{{\boldsymbol \ell}'\in \N^{N+2}}  | {\boldsymbol \ell}' \rangle \langle {\boldsymbol \ell}' |,
\ee
the duality relation \eqref{dual:relation} is rewritten as
\be
\label{dual2:braket}
\langle {\bf m} | e^{-t \mathcal{H}^t} D | {\boldsymbol \ell} \rangle =   \langle {\bf m} | D  e^{-t\widetilde{\mathcal{H}}} | {\boldsymbol \ell} \rangle.
\ee
To prove this it is clearly enough to show that
\be
\label{dual3:braket}
\langle {\bf m} |  \mathcal{H}^t D | {\boldsymbol \ell} \rangle =   \langle {\bf m} | D  \widetilde{\mathcal{H}} |  {\boldsymbol \ell} \rangle.
\ee
Furthermore, to establish \eqref{dual3:braket}, considering the additive form of the
original Hamiltonian \eqref{eq:fullham} and the dual Hamiltonian
\eqref{eq:fullhamdual}, it is enough
to prove the single edge (self-)duality
\be
\label{single:edge:duality}
\langle {\bf m} |\mathcal{\mathcal{H}}^{t}_{i,i+1}  D|  {\boldsymbol \ell} \rangle  = \langle {\bf m} |D \mathcal{\mathcal{H}}_{i,i+1}|  {\boldsymbol \ell} \rangle,
\ee
and the boundary dualities
\be
\label{left:boundary:duality}
\langle {\bf m} |\mathcal{\mathcal{H}}^{t}_{1} D |  {\boldsymbol \ell} \rangle = \langle {\bf m} |D \widetilde{\mathcal{\mathcal{H}}}_{0,1}|  {\boldsymbol \ell} \rangle,
\ee
\be
\label{right:boundary:duality}
\langle {\bf m} |\mathcal{\mathcal{H}}^{t}_{N}  D |   {\boldsymbol \ell} \rangle = \langle {\bf m} |D \widetilde{\mathcal{\mathcal{H}}}_{N,N+1}|  {\boldsymbol \ell} \rangle.
\ee
These will be shown in the next two sections.

\subsubsection{Bulk duality}

We start by proving \eqref{single:edge:duality}.
To this aim we express 
the duality function in terms of the generators  
\eqref{eq:sl2action} of the $\mathfrak{sl}(2)$  algebra.
We recall \eqref{factorial} by which we may write
\begin{equation}
\binom{m_i}{\ell_i} = \langle m_i|e^{S_+^{[i]}}|\ell_i\rangle.
\end{equation} 
Thus, inserting the previous expressions into \eqref{duality:function} we find 
\be
\label{dual:fct}
D({\boldsymbol m} , {\boldsymbol\ell}) = \rho_a^{\ell_0} \cdot \Big[\prod_{i=1}^N \langle m_i|e^{S_+^{[i]}}|\ell_i\rangle\Big] \cdot \rho_b^{\ell_{N+1}}
\ee
Notice that the duality function has
a product structure. 
As a consequence, since  $\mathcal{H}_{i,i+1}$ only acts on $|\ell_i \rangle \otimes|\ell_{i+1}\rangle$,  
we find that \eqref{single:edge:duality} is equivalent to
\be
\label{duality:bulk}
 \mathcal{H}^{t}_{i,i+1} e^{S_+^{[i]}} e^{S_+^{[i+1]}} = e^{S_+^{[i]}} e^{S_+^{[i+1]}} \mathcal{H}_{i,i+1}.
\ee
We observe that the Hamiltonian density is symmetric, i.e. $\mathcal{H}^{t}_{i,i+1} =  \mathcal{H}_{i,i+1}$.
Moreover, due to the $\mathfrak{sl}(2)$ symmetry of the Hamiltonian density (cf. \eqref{ham:fadeev}), we have
\be
\label{sl2:invariance}
[\mathcal{H}_{i,i+1}, S_+^{[i]} + S_+^{[i+1]}] = 0.
\ee
This can also be verified by a simple explicit computation
that just uses the definition \eqref{eq:harmonicaction} of the Hamiltonian density 
and the definition \eqref{eq:sl2action} of the creation operator. 
As a result,  \eqref{sl2:invariance} implies \eqref{duality:bulk}, which in turn
proves \eqref{single:edge:duality}.

We remark that the duality relation \eqref{single:edge:duality} explicitly reads:
\begin{eqnarray}
\label{no:trivial:id}
& &\sum_{k=1}^{m_i} \frac{1}{k} \Big\{ {m_i - k \choose {\ell}_i} {m_{i+1} + k \choose {\ell}_{i+1}} - {m_i  \choose {\ell}_i} {m_{i+1}  \choose {\ell}_{i+1}} \Big \} + 
\nonumber\\
& &
\sum_{k=1}^{m_{i+1}} \frac{1}{k} \Big\{ {m_i + k \choose {\ell}_i} {m_{i+1} - k \choose {\ell}_{i+1}} - {m_i  \choose {\ell}_i} {m_{i+1}  \choose {\ell}_{i+1}} \Big \} =
\nonumber\\
& &
\sum_{k=1}^{{\ell}_i} \frac{1}{k} \Big\{ {m_i  \choose {\ell}_i -k} {m_{i+1}  \choose {\ell}_{i+1}+k} - {m_i  \choose {\ell}_i} {m_{i+1}  \choose {\ell}_{i+1}} \Big \} + 
\nonumber\\
& &
\sum_{k=1}^{{\ell}_{i+1}} \frac{1}{k} \Big\{ {m_i  \choose {\ell}_i +k} {m_{i+1} \choose {\ell}_{i+1}-k} - {m_i  \choose {\ell}_i} {m_{i+1}  \choose {\ell}_{i+1}} \Big \}. 
\end{eqnarray}
A direct proof of this relation seems very difficult to obtain. Changing perspective, we see that the 
$\mathfrak{sl}(2)$ symmetry can be used to deduce the non trivial identity \eqref{no:trivial:id}.

\subsubsection{Boundary duality}

We now prove the right boundary duality \eqref{right:boundary:duality}.
Since the action of ${\cal H}_N$ and $\widetilde{{\cal H}}_{N,N+1}$
is local,  \eqref{right:boundary:duality} is equivalent to
\begin{eqnarray}
\label{no:trivial:id:boundary}
& &\sum_{k=1}^{m_1} \frac{1}{k} \Big\{  {m_{N} - k \choose {\ell}_{N}} -  {m_{N}  \choose {\ell}_{1}} \Big \} + 
\sum_{k=1}^{\infty} \frac{\beta_N^k}{k} \Big\{  {m_N + k \choose {\ell}_N} -   {m_N  \choose {\ell}_1} \Big \} =
\nonumber\\
& &
\sum_{k=1}^{{\ell}_1} \frac{1}{k} \Big\{ \rho_b^{k}{m_N  \choose {\ell}_N-k}  -  {m_N  \choose {\ell}_N}  \Big \}. 
\end{eqnarray}
%
One way to show that the relation in \eqref{no:trivial:id:boundary} holds is using the boundary Yang-Baxter equation. We have discussed in Section~\ref{sec:baegen} how the K-matrices \eqref{eq:kfunds} can be brought to an upper triangular form by a similarity transformation, cf.~\eqref{eq:u1} and \eqref{eq:u2}. 
%
Inserting the similarity transformation in \eqref{eq:u1} into the boundary Yang-Baxter equation \eqref{eq:bybe1} under the identification of the parameters \eqref{eq:identify} yields
\begin{equation}\label{eq:bybe1mod}
 L(x-y)e^{S_-}\hat{\mathcal{K}}(x)e^{-S_-}L(x+y)\hat K'''(y)=\hat K'''(y)L(x+y)e^{S_-}\hat{\mathcal{K}}(x)e^{-S_-}L(x-y)\,.
\end{equation} 
Alternatively we can solve again the boundary Yang-Baxter equation with  $K''$ in the two-dimensional space and read off the solution from \eqref{eq:offK} by specifying the parameters appropriately. In this way we derive the relation
\begin{equation}\label{eq:kkkk}
 e^{S_-}\hat{\mathcal{K}}(x)e^{-S_-}=f(x)e^{\frac{\beta}{1-\beta}S_+}\frac{\Gamma(1-x)}{\Gamma(1+x)}\frac{\Gamma(\frac{1}{2}+S_0+x)}{\Gamma(\frac{1}{2}+S_0-x)}e^{-\frac{\beta}{1-\beta}S_+}\,.
\end{equation} 
Here $f(x)$ is a normalisation that cannot be determined by the boundary Yang-Baxter equation, cf.~\eqref{eq:diagK}.
We fix the normalisation in \eqref{eq:kkkk} by comparing both sides for $\beta=0$ assuming that $f(x)$ does not depend on $\beta$. We find that $f(x)=1$.
The relation \eqref{eq:kkkk} for the original K-matrix $\hat{\mathcal{K}}(x)$ and the triangular K-matrix $e^{S_-}\hat{\mathcal{K}}(x)e^{-S_-}$ then immediately yields a relation for the boundary term of the Hamiltonian. 
By taking the logarithmic derivative at $x=0$, we find that the Hamiltonian ${\cal H}_N$ enjoys the property
\begin{equation}\label{eq:bndtrr}
 e^{S_-^{[N]}}\mathcal{H}_{N}e^{-S_-^{[N]}}=e^{\frac{\beta_N}{1-\beta_N} S_+^{[N]}}\psi(S_0^{[N]}+\frac{1}{2})e^{-\frac{\beta_N}{1-\beta_N} S_+^{[N]}}-\psi(1)\,.
\end{equation} 
Thus, introducing the boundary operator $\widetilde{{\cal H}}_N$
\begin{equation}\label{eq:absorb}
\widetilde{{\cal H}}_N=e^{-\frac{\beta_N}{1-\beta_N} S_-^{[N]}}\psi(S_0^{[N]}+\frac{1}{2})e^{\frac{\beta_N}{1-\beta_N} S_-^{[N]}}-\psi(1)\,,
\end{equation} 
we obtain the identity
\begin{equation}\label{eq:bndtr}
 e^{S_-^{[N]}}\mathcal{H}_{N}e^{-S_-^{[N]}}=\left(\widetilde{{\cal H}}_N\right)^t\,.
\end{equation} 
In other words, $e^{-S_-}$ is the intertwiner between $\mathcal{H}_{N}$ and $\left(\widetilde{{\cal H}}_N\right)^t$.
Now it is straightforward to show \eqref{no:trivial:id:boundary} using the relation \eqref{eq:bndtr}. 
Indeed, from the computation done in Section~\ref{sec:rbd}, we find
\begin{equation}
\begin{split}
 &\quad\, \bigg(\langle \ell_N | e^{S_-^{[N]}}\mathcal{H}_N|m_N\rangle \bigg)^t\\&=
 \left(\langle \ell_N | e^{S_-^{[N]}}\left(\left(h(m_N)+\sum_{k=1}^\infty\frac{\beta^k_N}{k}\right)|m_N\rangle -\sum_{k=1}^{m_N} \frac{1}{k}|m_N-k\rangle -\sum_{k=1}^\infty \frac{\beta^k_N}{k}|m_N+k\rangle \right)\right)^t\\
 &=
\sum_{k=1}^{m_N} \frac{1}{k} \Big\{ {m_N  \choose \ell_N}-{m_N - k \choose \ell_N}  \Big \} + 
\sum_{k=1}^{\infty} \frac{\beta^k_N}{k} \Big\{    {m_N  \choose \ell_N} -{m_N + k \choose \ell_N} \Big \} \,,
\end{split}
\end{equation} 
and
\begin{equation}
\begin{split}
  \left(\langle \ell_N | \left(\widetilde{{\cal H}}_N\right)^te^{S_-^{[N]}}|m_N\rangle \right)^t=
 \langle m_N| e^{S_+^{[N]}}\widetilde{{\cal H}}_N|\ell_N\rangle&=\langle m_N| e^{S_+^{[N]}}\left(h(\ell_N)|\ell_N\rangle-\sum_{k=1}^{\ell_N}\frac{1}{k}|\ell_N-k\rangle\right)=\\
 &=
\sum_{k=1}^{\ell_N} \frac{1}{k} \Big\{   {m_N  \choose \ell_N}-\rho^{k}_a{m_N  \choose \ell_N -k}   \Big \}\,,
\end{split}
\end{equation} 
with $\rho_a=\beta_N/(1-\beta_N)$. The identity
\eqref{eq:bndtr} shows the relation \eqref{no:trivial:id:boundary}
and thus the right boundary duality \eqref{right:boundary:duality}. 
The proof of the  left boundary duality \eqref{left:boundary:duality} is similar and is omitted.

\subsection{Correlation functions}
\label{dual-consequences}

In this section we show that duality allows to study the correlation functions
of the boundary driven process in terms of the dynamics of $n$ dual walkers. 

For $1\le i_1 \le i_2 \le \ldots \le i_n \le N$,
the $n$-point correlation functions of $\{{\bf M}(t)\, , \, t\ge0\}$ are defined by
\be
\label{correlation:fct}
\mathbb{E} [M_{i_1}(t)M_{i_2}(t)\ldots M_{i_n}(t)]
= 
\sum_{{\bf m} \in \mathbb{N}^N} m_{i_1} m_{i_2}\ldots m_{i_n} \mu_t({\bf m})
\ee
where $\mu_t({\bf m}) = \mathbb{P}({\bf M}(t) = {\bf m})$ is the law of the 
process at time $t\ge 0$. 
In particular we shall be interested in the $n$-point correlation functions 
in the stationary invariant state of the dynamics that is reached in the
limit of very large times, i.e.
\be
\mathbb{E} [M_{i_1}M_{i_2}\ldots M_{i_n}]
= 
\lim_{t\to\infty} \mathbb{E} [M_{i_1}(t)M_{i_2}(t)\ldots M_{i_n}(t)]
\ee 
The product $M_{i_1}(t)M_{i_2}(t)\ldots M_{i_n}(t)$ can in general be
obtained from the duality function in \eqref{duality:function}
evaluated in special dual configurations.
For instance, if the indexes are all different,  i.e. $1\le i_1 < i_2 < \ldots < i_n \le N$,
then, choosing the special dual configuration ${\boldsymbol \ell}^{\star}$ defined by
\be
 {\boldsymbol \ell}^{\star}_i =  
 \left\{\begin{array}{rl}
 1 & \text{if } x \in\{i_1,\ldots,i_n\},\\
 0 & \text{otherwise}.
 \end{array} \right.
\ee
it follows from \eqref{duality:function} that 
\be
D({\bf M(t)},  {\boldsymbol \ell}^{\star}) = M_{i_1}(t)M_{i_2}(t)\ldots M_{i_n}(t).
\ee
As a consequence of the duality relation \eqref{dual:relation},
one can express the $n$-point correlation functions of the original process 
defined by the Hamiltonian \eqref{eq:fullham} in terms
of the dynamics of the dual process (defined by the  Hamiltonian 
\eqref{eq:fullhamdual}) initialized with $n\in\mathbb{N}$ dual particles. 
Thus, by duality, we can study the
original systems by looking at a dual system with {\em finitely many particles}. 
In particular, since the sites $0$ and $N+1$ are  absorbing for the dual particles,
we can express the expectation of the duality function in the stationary state of the original 
system in terms of the {\em absorption probabilities} of the dual walkers.

To show this we take the limit $t\to\infty$ in \eqref{dual:relation}.
We first consider the left hand side of \eqref{dual:relation}, where
the evolution concerns the  process $\{{\bf M}(t)\, , \, t\ge0\}$ . 
In \eqref{dual:relation}, the initial configuration of the  process is  
chosen to be ${\bf m}=(m_1,\ldots,m_N) \in \mathbb{N}^{N}$, 
thus the initial distribution is $|{\mu}_0 \rangle = | {\bf m} \rangle$. 
The law of the process at time $t\ge 0$ is described by the 
ket $|{\mu}_t \rangle$, that encodes the time-dependent distribution ${\mu}_t({\bf m})$ as
\be
|\mu_t\rangle = \sum_{{\bf m}\in\mathbb{N}^N} \mu_t({\bf m})  |{\bf m}\rangle  . 
\ee
and solves the master equation, which in bra-ket notation is written as the Schr\"odinger equation
with Euclidean time
\be
\label{master}
\frac{d}{dt}|\mu_t\rangle = -{\cal H} |\mu_t\rangle.
\ee
In the limit of large times, regardless of the choice of the initial configuration ${\bf m}$, the  process will 
approach its non-equilibrium  steady state, i.e.
\be
\lim_{t\to\infty} e^{-t\mathcal{H}} | {\bf m} \rangle =  | \mu \rangle
\ee
where  $| \mu \rangle$ is the stationary solution of \eqref{master}, i.e.
it solves ${\cal H} |\mu \rangle = 0$. 
As a consequence, for left hand side of \eqref{dual:relation} we have
\begin{eqnarray}
\label{leftdual}
\lim_{t\to\infty} \mathbb{E}_{{\bf m}} \Big[D({\bf M}(t), {\boldsymbol \ell}) \Big]
& = &
\lim_{t\to\infty}  \sum_{{\bf m}'\in \N^{N}} \langle {\bf m}' | D | {\boldsymbol \ell}\rangle \langle {\bf m}' | e^{-t \mathcal{H}} | {\bf m} \rangle 
\nonumber \\
& = & 
\sum_{{\bf m}'\in \N^{N}} \langle {\bf m}' | D | {\boldsymbol \ell} \rangle \langle {\bf m}' |  \mu \rangle
\nonumber \\
& = & 
\sum_{{\bf m}'\in \N^{N}}  D({\bf m}', {\boldsymbol \ell})   \mu({\bf m}').
\end{eqnarray}

We now move to the right hand side of \eqref{dual:relation}, where
the evolution concerns the dual process $\{{\bf L}(t) \, , t\ge 0\}$. 
We recall that the dual process {\em conserves the total number
of  particles} and it has two absorbing sites. 
Thus, if initially the dual process  is started
from the measure $|\widetilde{\mu}_0 \rangle = | {\boldsymbol \ell} \rangle$
concentrated on the initial configuration ${\boldsymbol \ell}=(\ell_0,\ell_1,\ldots,\ell_N,\ell_{N+1}) \in \mathbb{N}^{N+2}$,
in the long time limit it will approach its stationary distribution 
\be
\lim_{t\to\infty} e^{-t\widetilde{\mathcal{H}}} | {\boldsymbol \ell} \rangle =  | \widetilde{\mu}^{({\boldsymbol\ell})} \rangle.
\ee
At variance with the original process, the stationary state of the dual process
will depend on the initial configuration ${\boldsymbol \ell}$, more precisely it will depend
on the total number of particles $|{\boldsymbol \ell}| = \sum_{i=0}^{N+1} \ell_i$. Furthermore,
given the properties
of the dual dynamics, the stationary state will concentrate
on the subset of configurations
\be
\Omega_{{\boldsymbol \ell},N} = \{{\boldsymbol \ell}' \in \mathbb{N}^{N+2} \; : \; {\boldsymbol \ell}'=(l_0',0,\ldots,0,{\ell}'_{N+1}), \;  \; {\ell}'_0 + {\ell}'_{N+1} = |{\boldsymbol \ell}| \}.
\ee 
Thus
\be
|\widetilde{\mu}^{({\boldsymbol \ell})}\rangle = \sum_{{\boldsymbol \ell}'\in\Omega_{{\boldsymbol \ell},N}} \widetilde{\mu}^{({\boldsymbol \ell})}({\boldsymbol \ell}')  |{\boldsymbol \ell}'\rangle.
\ee
Equivalently we can write
\be
|\widetilde{\mu}^{({\boldsymbol \ell})}\rangle = \sum^{\quad *}_{{\ell}'_0,{\ell}'_{N+1}} {q}_{\boldsymbol \ell}({\ell}'_0,{\ell}'_{N+1})  \, |{\ell}'_0,0,\ldots,0,{\ell}'_{N+1}\rangle.
\ee
where $\sum^*$ denotes the sum restricted to $({\ell}'_0,{\ell}'_{N+1}) \in \mathbb{N}^2$ such that
${\ell}'_0 + {\ell}'_{N+1} = |{\boldsymbol \ell}|$ and ${q}_{\boldsymbol \ell}({\ell}'_0,{\ell}'_{N+1})$ denotes the probability that,
being the dual particles initially placed as prescribed by the
configuration ${\boldsymbol \ell}$, eventually ${\ell}'_0$ of them are absorbed in $0$
and the remaining ${\ell}'_{N+1} = |{\boldsymbol \ell}| - {\ell}'_0$ are absorbed 
in $N+1$. 

If we use all this in the right hand side of \eqref{dual:relation} we find
\begin{eqnarray}
\label{rightdual}
\lim_{t\to\infty}  \widetilde{\mathbb{E}}_{{\boldsymbol \ell}} \Big[D({\bf m}, {\bf L}(t))\Big]
& = &
\lim_{t\to\infty}  \sum_{{\boldsymbol \ell}'\in \N^{N+2}} \langle {\bf m} | D | {\boldsymbol \ell}' \rangle \langle {\boldsymbol \ell}' | e^{-t\widetilde{\mathcal{H}}} | {\boldsymbol \ell} \rangle
\nonumber \\
& = & 
\sum_{{\boldsymbol \ell}'\in \N^{N+2}} \langle {\bf m} | D | {\boldsymbol \ell}' \rangle \langle {\boldsymbol \ell}' | \widetilde{\mu}^{({\boldsymbol \ell})}  \rangle
\nonumber \\
& = & 
\sum^{\quad *}_{({\ell}'_0,{\ell}'_{N+1})\in\N^2}  \rho_a^{{\boldsymbol \ell}'_0} \; \rho_b^{{\boldsymbol \ell}'_{N+1}}  \; {q}_{\boldsymbol \ell}({\ell}'_0,{\ell}'_{N+1}),
\end{eqnarray}
where in the last equality it has been used that
\be
\langle m_1,\ldots,m_N | D | {\ell}'_0,0,\ldots,0,{\ell}'_{N+1}\rangle = \rho_a^{{\ell}'_0} \; \rho_b^{{\ell}'_{N+1}},
\ee
which follows from the definition of the duality function \eqref{duality:function}.

Taking the limit $t\to\infty$ in \eqref{dual:relation} and using \eqref{leftdual} and \eqref{rightdual} 
we arrive to the main result of this section:
\be
\label{dual-formula}
\lim_{t\to\infty} \mathbb{E}_{\bf m}[D({\bf M}(t), {\boldsymbol \ell})] = \sum^{\quad *}_{({\ell}'_0,{\ell}'_{N+1})\in\N^2}  \rho_a^{{\ell}'_0} \; \rho_b^{{\ell}'_{N+1}}  \; {q}_{\boldsymbol \ell}({\ell}'_0,{\ell}'_{N+1}).
\ee
We close by showing the applications of this formula to the one and two points correlation function.

\paragraph{One point correlation function.} We take  $n=1$ in \eqref{correlation:fct} and put $i_1 =i$. 
Considering the dual configurations ${\boldsymbol \delta}_{i}$ that have one particle
at site $i$ and zero elsewhere, i.e. 
\be
({\boldsymbol \delta}_i)_k =  
 \left\{\begin{array}{rl}
 1 & \text{if } k=i,\\
 0 & \text{otherwise},
 \end{array} \right.
\ee
from  \eqref{duality:function} we have  that
\begin{eqnarray}
&&D({\bf M}(t),{\boldsymbol \delta}_0)  =  \rho_a  \nonumber \\
&&D({\bf M}(t),{\boldsymbol \delta}_i)  =  M_i(t) \qquad i=1,\ldots, N \nonumber \\ 
&&D({\bf M}(t),{\boldsymbol \delta}_{N+1})  =  \rho_b  
\end{eqnarray}
For the average number of particles at site $i$, being the process
 started from the measure $\mu$, we have 
\begin{eqnarray}
\mathbb{E}(M_i(t)) 
& = & 
\int d\mu({\bf m}) \mathbb{E}_{\bf m}[D({\bf M}(t), {\boldsymbol \delta}_i)] \nonumber \\
& = & 
\int d\mu({\bf m}) \widetilde{\mathbb{E}}_{\bf {\boldsymbol \delta}_i}[D({\bf m}, {\boldsymbol \delta}_i(t))] 
\end{eqnarray}
where in the last equality we used duality \eqref{dual:relation}. 
Since the dual process has one particle only,
it follows from the dual Hamiltonian \eqref{eq:fullhamdual} 
that this particle moves
as a continuous time symmetric random walk 
$\{I(t)\,,\,t\ge0\}$ started at $i\in\{1,\ldots,N\}$,  jumping at rate 1 on $\{0,1,\ldots,N,N+1\}$ 
and being absorbed at $\{0\}$ and at $\{N+1\}$.
Thus
\begin{eqnarray}
\mathbb{E}(M_i(t)) 
& = & 
\int d\mu({\bf m}) \mathbb{E}_{\bf {\boldsymbol \delta}_i}[D({\bf m}, {\boldsymbol \delta}_{I(t)})] .
\end{eqnarray}
By taking the limit $t\to\infty$ of the above expression, formula \eqref{dual-formula} allows to  compute the  profile
in the non-equilibrium stationary state:
\begin{eqnarray}
\mathbb{E}(M_i)&= & \lim_{t\to\infty} \mathbb{E}(M_i(t))  = \rho_a \, {q}_{{\boldsymbol \delta}_{i}}(1,0) + \rho_b \, {q}_{{\boldsymbol \delta}_{i}}(0,1).
\end{eqnarray}
The absorption probabilities of the simple symmetric random walk $\{I(t)\,,\,t\ge0\}$ are easily computed thus implying
\begin{eqnarray}
\mathbb{E}(M_i)
&=& \rho_a \Big(1-\frac{i}{N+1}\Big) + \rho_b  \Big(\frac{i}{N+1}\Big)\nonumber\\
&=& \rho_a + \frac{\rho_b-\rho_a}{N+1} i.
\end{eqnarray}

\paragraph{Two point correlation function.} 
We take  $n=2$ in \eqref{correlation:fct},  put $i_1=i$ and $i_2=j$.
To analyze the two point function $\mathbb{E}[M_i(t) M_j(t)]$ with $i\neq j$
we need to run the dual process with two particles initially at sites $i$ and $j$.
We can read from the dual Hamiltonian \eqref{eq:fullhamdual}
that they move as  two random walks $\{(I(t),J(t))\,,\,t \ge 0\}$
on $\{0,\ldots,N+1\}^2$ that are absorbed at $0$ and at $N+1$ 
and evolve via the generator
\begin{eqnarray}
L^{dual,2}f(i,j) & = & \Big[f(i-1,j) + f(i+1,j) + f(i,j-1) + f(i,j+1) - 4f(i,j)\Big] \mathbf{1}_{\{i\neq j\}} \nonumber \\
& + & \Big[(f(i-1,i) -f(i,i)) + \frac12 (f(i-1,i-1) -f(i,i)) \nonumber \\
& & + (f(i,i+1) -f(i,i)) + \frac12 (f(i+1,i+1) -f(i,i))\Big] \mathbf{1}_{\{i= j\}}\nonumber
\end{eqnarray}
with 
$$
L^{dual,2} f(0,0) =  L^{dual,2} f(0,N+1) =  L^{dual,2} f(N+1,0)  =  L^{dual,2} f(N+1,N+1) = 0.
$$
Duality \eqref{dual:relation} yields
\begin{eqnarray}
\mathbb{E}(M_i(t)M_j(t)) 
& = & 
\int d\mu({\bf m}) \mathbb{E}_{\bf {\boldsymbol \delta}_i+ {\boldsymbol \delta}_j}[D({\bf m}, {\boldsymbol \delta}_{I(t)}+{\boldsymbol \delta}_{J(t)})] .
\end{eqnarray}
Formula \eqref{dual-formula} tell that the two point functions in the stationary state 
can be written in terms of the absorption probabilities of the dual walkers as
\begin{eqnarray}
\mathbb{E}(M_i M_j)&= & \lim_{t\to\infty}\mathbb{E}[M_i(t)M_j(t)]  \nonumber \\
&= & \rho_a^2 \, {q}_{{\boldsymbol \delta}_{i}+ {\boldsymbol \delta}_{j}}(2,0) + \rho_a\rho_b \, {q}_{{\boldsymbol \delta}_{i}+{\boldsymbol \delta}_{j}}(1,1) + 
\rho_b^2 \, {q}_{{\boldsymbol \delta}_{i}+ {\boldsymbol \delta}_{j}}(0,2).
\end{eqnarray}
It would be interesting to investigate the application of Bethe ansatz to compute the 
absorption probabilities.

\subsection{Other dualities}
The duality results discussed so far can be generalized into several directions.
Firstly,  the process $\{{\bf L}(t) \, ,\, t\ge 0\}$ with Hamiltonian \eqref{eq:fullhamdual},
besides being the dual process of  $\{{\bf M}(t) \,,\, t\ge 0\}$ with Hamiltonian \eqref{eq:fullham},
is also the dual process of $\{{\bf X}(t) \,,\, t\ge 0\}$ with generator \eqref{ham:levy}.
It can be verified with an explicit computations that the duality
functions now read
\be
 \label{duality:function2}
D({\bf x}, {\boldsymbol \ell}) = \lambda_1^{\ell_0} \cdot \Big[\prod_{i=1}^N   \frac{{x_i}^{\ell_i}}{\ell_i!}\Big ] \cdot \lambda_N^{\ell_{N+1}}
\ee
This is ax example of the situation -- that was alluded to in the introduction -- where a transport model
 of a continuous quantity, interpreted as energy, has a dual that instead transport discrete particles. 
Algebraically, this is a consequence of a change of representation for the same
abstract Hamiltonian, that we identified in \eqref{ham:fadeev}. The results on the 
correlations described in Section \ref{dual-consequences} hold true mutatis mutandis,
now for the $n$-point correlation function of the energies
$
\mathbb{E} [X_{i_1}(t)X_{i_2}(t)\ldots X_{i_n}(t)].
$

A second generalization is the one to higher spin values. We discuss here only the closed periodic chain.
Similarly to what has been done in Section \ref{dual-proof}, one can show that 
the process $\{{\bf M}^{(s)}(t) \, ,\, t\ge 0\}$ with Hamiltonian \eqref{Hs} is self-dual with self-duality function
\be
D({\bf m}, {\boldsymbol \ell}) = \prod_{i=1}^N   \frac{m_i!}{(m_i-\ell_i)!} \frac{\Gamma(2s)}{\Gamma(\ell_i+2s)}.
\ee

Finally, we may also expect that orthogonal duality functions arise when considering 
unitary equivalent representations of the $\mathfrak{sl}(2)$ algebra, as it happens 
in the case of the spin chain related to the KMP model \cite{franceschini2017stochastic,franceschini2018self,carinci2018orthogonal}.

\section{Connection with $\mathcal{N}=4$ SYM}
\label{sec:4sym}

\subsection{Semiclassical string and fluctuating hydrodynamics}

As mentioned in the introduction, the imaginary and real time versions of the same $\mathfrak{sl}(2)$ chain are the stochastic
and the quantum models, respectively. The latter is derived from expectation values of local operators in $\mathcal{N}=4$ SYM theory \cite{Minahan:2002ve,Beisert:2003yb,Beisert:2003jj}, and its
long wavelength limit is described by a string spinning on $AdS_5\times S^5$  \cite{bellucci-sl2,Stefanski:2004cw}, see also \cite{Tseytlin:2004xa,Plefka:2005bk,Beisert:2010jr} for an overview. 
One may obtain in a similar manner the coarse-grained dynamics of the stochastic model -- its Fluctuating
Hydrodynamics -- and it turns out to be the Euclidean counterpart of the string equation.

Fluctuating Hydrodynamics is a stochastic process, and as such  has an associated supersymmetry which encodes,
through its Ward-Takahashi identities, probability conservation and the equilibrium theorems: time-translational invariance and 
fluctuation-dissipation relations \cite{Zinn,barci}. The supersymmetric extension of the stochastic chain is
 easy to construct directly  by exponentiating a Jacobian with fermions.
We show here that, interestingly, this supersymmetric chain is equivalent to the  $\mathfrak{sl}(2|1)$ superstring obtainable
directly from $\mathcal{N}=4$ SYM, which has been discussed in the literature \cite{bellucci-super}.  In other words, we follow the
`stochastic quantization' procedure (see Ref. \cite{dijkgraaf} for an overview of this) to obtain the chain and its susy completion.

Here we follow \cite{kruczenski,bellucci-sl2} and \cite{Tailleur}.
The derivation  of the string limit of the quantum system and of the hydrodynamic limit of the stochastic system  are  the same, for real and Euclidean times, respectively. 
We choose the following definition for the coherent state:
\begin{equation}\label{coh-st:gen}
  |z \rangle \propto e^{z S_+}\, |{0}\rangle \qquad\qquad
\end{equation}
where $|{0}\rangle$ is in the spin $s$ representation and is such that $S_0|{0}\rangle=s|{0}\rangle$.
 We have the expectation values (the so-called Q-symbols of the operators)
 \begin{eqnarray}
  \label{eqn:spinopsu11}
  \langle z | S_+ | z\rangle= 2s\frac { \bar z}{1-z \bar z} &\equiv& s \; n_+\nonumber \\
  \langle z | S_- | z\rangle =2s \frac{z}{1-z \bar z} &\equiv& s \;  n_- \nonumber\\
  \langle z | S_0 | z\rangle = s\frac{1+z \bar z}{1-z\bar z}&\equiv&s \;  n_0
\end{eqnarray}
where $\vec n$ spans the hyperboloid:
 \[
\vec{n}^2=  n_0^2-n_1^2-n_2^2=1,\qquad n_0>0~.
\]

The semiclassical limit is achieved,  even for a single site, in the large `spin' ($s$) limit. In this limit, 
operators may be replaced by their expectations (the $Q$ symbols), and commutators by Poisson brackets.
Here (and in references \cite{kruczenski,bellucci-sl2,Tailleur}) we are not quite interested in the large spin $s$ limit, but in
a coarse-grained (long wavelength) description.  Making the implicit assumption that the hydrodynamic limit does not
depend on the spin, one first does a semiclassical large-$s$ approximation and then a long-wavelength approximation. In the probabilistic literature there are, however, direct derivations of fluctuating hydrodynamics without assuming large spins \cite{jona}.
In the semiclassical limit,  commutators become Poisson brackets
\begin{equation}
\{n_0,n_\pm\}=\pm n_\pm \;   \;\;\;\; ; \;\;\;\; 
\{n_+,n_-\}=-2n_0 \; 
\label{poisson}
\end{equation}

Let us also recall the
action of the spin operators in the coherent state representation:
\begin{eqnarray}
  \langle z |S_+ |\psi\rangle &=& \left[2s \bar z  + \bar z ^2
    \frac\partial{\partial \bar z }\right] \langle z  | \psi \rangle = 
    s \left[2  \hat  {\mathbf \rho} +  \hat  {\mathbf \rho}^2 \rho\right] \langle z  | \psi \rangle
  \nonumber\\
   \langle z |S_- |\psi\rangle &=& \frac\partial  {\partial \bar z } \langle z |\psi\rangle 
   = s \;  {\mathbf \rho}\langle z |\psi\rangle
              \nonumber
              \\ \langle z |S_0 |\psi\rangle &=& \left[\bar z  \frac{\partial}{\partial \bar z }+s \right] \langle z|\psi\rangle
    =s \left[ \hat  {\mathbf \rho} \rho+1 \right] \langle z|\psi\rangle          
              \label{eqn:SkAction}
\end{eqnarray}
yielding the Bargmann-Fock representation. Here we have defined
\begin{equation}
  {\mathbf \rho}= \frac 1 s \frac\partial{\partial \bar z } \;\;\;\; ; \;\;\;\; \bar z= \hat  {\mathbf \rho}\;\; \Rightarrow \;\; [\rho,\hat \rho]=\frac{1}{s}
 \end{equation}
Again, in the semiclassical limit we replace commutator by Poisson brackets $s \; [\;,\;] \; \rightarrow \; \{\;,\;\} $
\begin{eqnarray}
  n^+  &=& 
  2  \hat  { \rho} +  \hat  { \rho}^2 \rho
  \nonumber\\
  n^-  &=& 
    { \rho}
              \nonumber
              \\ n^0  &=&
     \hat  { \rho} \rho+1  
              \label{rhohatrho}
\end{eqnarray}
where here and in what follows $(\rho,\hat \rho)$ are c-numbers, and  (\ref{poisson}) becomes $\{ \rho, \hat \rho\}=1$.

We have two alternatives for describing the model in the semiclassical limit: through the $(n_\pm,n_0)$ or through $( \rho,\hat \rho)$ on each site, for the full chain through the variables:
\begin{eqnarray}
\{n_0(x),n_\pm(y)\}&=&\pm n_\pm(x) \; \delta(x-y)  \;\;\;\; ; \;\;\;\; 
\{n_+(x),n_-(y)\}=-2n_0 \; \delta(x-y) \nonumber \\
\{\rho(x),\hat\rho(y)\} &=& \; \delta(x-y)
\end{eqnarray}

The Hamiltonian of the chain  has a simple  (Q-symbol) representation in the
coherent state basis $|{\vec{n}}\rangle$  in \cite{Stefanski:2004cw,bellucci-sl2}:
\begin{eqnarray}\label{Ham:final}
   \langle{\vec{n}^k \vec{n}^{k+1}}| \, H^{k\,k+1}\, |{\vec{n}^k \vec{n}^{k+1}}\rangle
  &=& \log\left(1-\frac{\left(\vec{n}^k-\vec{n}^{k+1}\right)^2}{4}\right)~.
\end{eqnarray}
Here
$$\vec{n}^k\cdot\vec{n}^{k+1}=n_0^{k} n_0^{k+1}-n_+^{k} n_-^{k+1}-n_-^{k} n_+^{k+1}~.$$
In the 
hydrodynamic limit of large $k$ and tending to the continuum the Hamiltonian reads
\begin{equation}
\label{eqn:hamSu11}
{\cal{H}} = -\frac 1 2 \left[ \partial_x n_+ \cdot \partial_x n_--(\partial_x n_0)^2\right]
\end{equation}
 The same result in real time is found by considering semiclassical strings
 spinning on $AdS_5\times S^5$ \cite{bellucci-sl2}.
 
 In Euclidean time we obtain the hydrodynamic limit of the stochastic chains (see \cite{Tailleur}, where  this has been done for the 
hydrodynamic limit of the Kipnis-Marchioro-Presutti chain, but the result is the same).
To see this in detail,  we go to the $(\rho,\hat{\rho})$ variables (\ref{rhohatrho}):
  \begin{eqnarray}
{\cal H}(\vec n)&=& -\frac 12  \int dx \; \{\partial_x n_+ \partial_x n_- - (\partial_x n_0)^2 \}= -\frac 12 \int dx \; \{\partial_x(\hat \rho \rho + 2 \hat \rho) \; \partial_x \rho - [\partial_x(\hat \rho \rho)]^2\} \nonumber\\
&=&  \int dx \;\ \{ \partial _x \hat \rho \partial_x \rho - \frac 12  \rho^2 (\partial_x \hat {\rho})^2\} \label{Hrho}
\end{eqnarray}

\subsection{Stochastic dequantization}

The path integral may  then be written as:
\begin{eqnarray}
& & \int D\rho D \hat \rho \;  e^{N \int dxdt\; \{- \hat \rho \dot \rho -  \partial _x \hat \rho \partial_x \rho + \frac 1 2  \rho^2 (\partial_x \hat \rho)^2\} }
\nonumber\\
&=&\int D\rho  D\xi D \hat \rho \; e^{N \int dxdt\; \{-\hat \rho \dot \rho -  \partial _x \hat \rho \partial_x \rho + \rho (\partial_x \hat \rho)\xi(x,t) -\frac 12 \xi(x,t)^2 \}}
\nonumber\\
&=&\int D\rho  D\xi \; \delta \left\{    \dot \rho(x,t)  - \partial_x [ \partial_x \rho(x,t)  - \rho \xi(x,t) ]\right\}       \; e^{-\frac 1 2\int dtdx\; \xi^2(x,t)   } 
\label{ll1}   \end{eqnarray}
where we have used Gaussian (Hubbard-Stratonovich)  decoupling with the field $\xi(x,t)$, integration by parts, and
 finally integrated over $\hat \rho$.
 The meaning of (\ref{ll1})  is clear, we have a stochastic system satisfying
\begin{equation}
{\mbox{Eq}}\left[\rho(x,t) \right] = 
\dot \rho(x,t)  - \partial_x [ \partial_x \rho(x,t)  - \rho(x,t) \xi(x,t) ]=0
\label{langevin}
\end{equation}
with Gaussian noise $\xi(x,t)$, white in space and in time. This is a hydrodynamics with fluctuating current $J= -\partial_x \rho(x,t)  + \rho(x,t) \xi(x,t) $.  
 We have been  careless about time-discretization of $\xi(x,t)$: this is abundantly discussed in the
literature (see \cite{barci}) of multiplicative noise.

The equations of motion for Euclidean (imaginary) time given by (\ref{Hrho}) are the Freidlin-Wenzel / WKB \cite{jona}  description
of the coarse-grained dynamics of the stochastic model  (\ref{langevin}), with $\rho(x,t)$ the density field and $\hat{\rho} (x,t)$ the `response fields'
familiar  \cite{six} in stochastic systems: 
\begin{equation}
\dot n_a(x,t) = \{ n_a, {\cal{H}}(\vec n) \}\;\; 
\label{neq}
\end{equation}
and 
\begin{equation}
\dot \rho(x,t) = \{ \rho, {\cal{H}}(\rho,\hat \rho) \}\;\; \;\; ; \;\;\;\; \dot {\hat \rho}(x,t) = \{ \hat \rho, {\cal{H}}(\rho,\hat \rho) \}
\label{req}
\end{equation}
Note that if $\rho$ and $\hat \rho$ are real, the equation (\ref{neq}) implies complex components for the $n_0,n_1,n_2$,
and vice-versa (in this sense, the equation for $\rho$, $\hat \rho$ is an instanton equation for the quantum model).
The semiclassical real-time equations are the same, with $t \rightarrow i\tau$.

\subsection{Stochastic requantization and superstring}

As is well known  \cite{Zinn}, a stochastic system can be promoted into a system having a BRST symmetry
(that guarantees probability conservation), plus,
{\em only} when the system satisfies detailed balance, an extra supersymmetry with the interpretation of thermodynamic equilibrium -- so that for example, Fluctuation Dissipation relations appear as Ward identities. For multiplicative
noise this has been less discussed, see however Refs \cite{barci}.
The construction is as follows: we retrace our steps starting from (\ref{langevin}), but this time exponentiating  the 
Jacobian term with Grassmann variables \cite{dijkgraaf}:
\begin{eqnarray}
1&=&\int D\rho  D\xi \; \delta \left\{    \dot \rho(x,t)  - \partial_x [ \partial_x \rho(x,t)  -
 \rho \xi(x,t) ]\right\} 
 {\mbox{det}}  \left|   \frac{\delta{{\mbox{Eq}}}\left[\rho(x,t) \right] }{\delta \rho(x',t')}    \right|
   \; e^{-
\frac 1 2 \int dtdx\; \xi^2(x,t)}       \nonumber \\
&=&\int D\rho D \hat \rho D\xi \;  D\eta \; D\bar \eta\;e^{N\int dt dx \;  \left\{-\hat \rho(x,t)\dot 
\rho(x,t)  + \hat \rho(x,t)\partial_x [ \partial_x \rho(x,t)  - \rho \xi(x,t)]-\frac 12  \xi^2(x,t)\right\}} \nonumber \\
& & \hspace{2cm}e^{-N\int dt dx dt' dx' \; \bar \eta(t,x)
 \frac{\delta{{ \mbox{Eq}}}\left[\rho(x,t) \right] }{\delta \rho(x',t')}    \eta(x', t')}\label{grass}\\
&=&\int D\rho D \hat \rho D\xi \;  D\eta \; D\bar \eta\;e^{N\int dt dx \; 
\left\{ -\hat \rho(x,t)\dot \rho(x,t) - \bar \eta (x,t)\dot \eta (x,t) + 
\hat \rho(x,t) \partial_x^2 \rho(x,t) \right\}}\nonumber \\
& & \hspace{2cm}e^{N \int dt dx \{
   [\rho(x,t) \partial_x \hat \rho{\color{black}+}  \partial_x \bar \eta(t,x)
 \eta(x, t)]  \xi(x,t)+ \partial_x^2 \bar \eta(t,x)  \eta(x, t)- \frac 12 \xi^2(x,t)
   \}}
\nonumber
\end{eqnarray}

 This corresponds to a system with fermions,
with the total number of them  $\langle \; \int dx \; \bar \eta(x) \eta(x)\rangle $ a conserved quantity. 
We may now integrate away the noise  variable $\xi$ to obtain:
\begin{eqnarray}
e^{-NS}&=&\int D\rho D \hat \rho  \;  D\eta \; D\bar \eta\;e^{ N\int dt dx \; \left\{-\hat \rho \; \dot \rho  - \bar \eta \;\dot \eta  - \hat \rho \; \partial_x^2 \rho  +
\frac 12 \rho^2 (\partial_x \hat \rho)^2 {\color{black} + }  \rho \; \partial_x  \hat \rho  \; \partial_x \bar   \eta \; \eta- \partial_x^2 \bar \eta \; \eta\right\}
  }\nonumber\\
  &=&\int D\rho D \hat \rho  \;  D\eta \; D\bar \eta\;e^{N \int dt dx \; \left\{-\dot {\hat \rho} \; \rho - \bar \eta \; \dot \eta  - \partial_x^2
\hat \rho  \; \rho  + \frac 12 
\rho^2 (\partial_x \hat \rho)^2 {\color{black} +}  \rho \partial_x  \hat \rho  \; \partial_x \bar \eta \; \eta+ \partial_x \bar \eta \; \partial_x \eta\right\}
  }
   \label{susy}
\end{eqnarray}
where we have integrated by parts in time and space, valid for a trace (periodic in time)  and a closed chain.  
The restriction to the zero-fermion subspace of  (\ref{susy}) yields back  (\ref{ll1}): this is the usual relation between supersymmetric quantum mechanics and stochastic processes \cite{Zinn}.
One has to be careful about the time-discretization of the stochastic process -- Ito, Stratonovich and other 
conventions, here it is encapsulated in the actual value equal-time expectations $\langle  \bar \eta(x,t) \eta(x,t)\rangle$ (see 
discussion in \cite{barci}).

Let us write an alternative expression for the action in (\ref{susy}). We make the change of variables $\rho(x,t) \rightarrow \rho(x,t) (1 - \frac 12 \bar \eta (x,t) \eta(x,t))$ 
\begin{eqnarray}
S &=& \int dt dx \; \left\{\dot {\hat \rho} \rho  +\bar \eta \; \dot \eta 
+ \partial_x^2 \hat \rho   \rho 
 - \frac 12  \rho^2 (\partial_x \hat \rho)^2 
 { -}   \rho \partial_x \hat \rho  \; \partial_x \bar \eta \; \eta 
 - \partial_x \bar \eta \; \partial_x \eta\right\} \nonumber \\
 &\rightarrow& \int dt dx \; \left\{\dot {\hat \rho} \rho  + \bar \eta \; \dot \eta - \frac 12 \dot {\hat \rho} \rho \; \bar \eta \; \eta
+ \partial_x^2 \hat \rho   \rho \; \left(1-\frac 12  \bar \eta \; \eta\right)          \right. \nonumber \\
& &  \left.    -\frac 12  \rho^2 (\partial_x \hat \rho)^2  \; (1- \bar \eta \; \eta)
 -   \rho \partial_x \hat \rho \;  \partial_x \bar \eta \;  \eta
 - \partial_x \bar \eta \; \partial_x \eta\right\} \nonumber \\
&=&  \int dt dx \; \left\{\dot {\hat \rho} \rho  + \bar \eta \; \left(\partial_t   -\frac 12  \dot {\hat \rho} \rho\right)        \;           \eta 
-\left[ \partial_x \hat \rho  \partial_x \rho  + \frac 12 \rho^2 (\partial_x \hat \rho)^2\right]\; \left(1-\frac 12  \bar {\eta} \; \eta\right)
 + \frac 12  \partial_x \hat \rho   \rho \; \partial_x ( \bar {\eta} \; \eta) \right. \nonumber \\
& & \;\;\;\;   \;  + \frac 14  \rho^2 (\partial_x \hat \rho)^2  \; \bar {\eta} \; \eta
\left. -   \rho \partial_x \hat \rho  \; \partial_x \bar \eta \;  \eta
 - \partial_x \bar \eta \; \partial_x \eta\right\} \nonumber \\
 &=&  \int dt dx \; \left\{\dot {\hat \rho} \rho  + \bar \eta \; \left(\partial_t   -\frac 12  \dot {\hat \rho} \rho\right)                   \eta 
-\left[ \partial_x \hat \rho  \partial_x \rho  + \frac 12 \rho^2 (\partial_x \hat \rho)^2\right]\; \left(1-\frac 12  \bar \eta \; \eta\right)   \right. \nonumber\\
& & \left.     
 + \frac 14  \rho^2 (\partial_x \hat \rho)^2  \; \bar {\eta} \; \eta + \frac 12  \partial_x \hat \rho   \rho \; \partial_x ( \bar \eta \; \eta) -                
        \rho \partial_x \hat \rho   \partial_x \bar \eta \;  \eta
 - \partial_x \bar \eta \; \partial_x \eta \right\} \nonumber  \\
&=&  \int dt dx \; \left\{\dot{\hat \rho} \rho  + \bar \eta \; \left(\partial_t   -\frac 12  \dot{\hat \rho} \rho\right)                   \eta 
-\left[ \partial_x \hat \rho  \partial_x \rho  + \frac 12  \rho^2 (\partial_x \hat \rho)^2\right]\; \left(1-\frac 12  \bar \eta \; \eta\right)\right. \nonumber \\
& & \;\;\;\;   \;  \left.
-\left(\partial_x + \frac 12 \rho \partial_x \hat \rho\right) \bar \eta \; \left(\partial_x - \frac 12  \rho \partial_x \hat \rho\right) \eta \right\}\nonumber\\
&=& \int dt dx \; \left\{\dot{\hat \rho} \rho  
+\left[ \partial_x \hat \rho  \partial_x \rho  + \frac 12 \rho^2 (\partial_x \hat \rho)^2\right]\; \left(1-\frac 12  \bar \eta \; \eta\right)
+ \bar \eta \; D_t^-                  \eta + D_x^+ \bar \eta \; D_x^- \eta \right\}
 \label{susy1}
\end{eqnarray}
where $D_\sigma^\pm \equiv \partial_\sigma \pm \frac 12 \rho \partial_\sigma \hat \rho$.

Written in this way, one may compare directly with the expression given in Ref. \cite{bellucci-super} for the  $\mathfrak{sl}(2|1)$ closed superstring.
 \footnote{ This may be seen directly by making the change of variables $\bar \eta \rightarrow -2\bar \xi $, $ \eta \rightarrow  \xi$
  for fermions, $\hat \rho \rightarrow 2\tilde \rho$ for the response field and $S \rightarrow 2S$ for the action. Note that our $\rho$ is not theirs, which is
  the `radial' coordinate of the hyperboloid.}

\section{Conclusion and outlook}
In this article we have studied the stochastic particle processes that arise from the non-compact Heisenberg chain. We explicitly gave the corresponding hopping rates for a spin chain with spin $s$ at each site and showed that they can be recovered from the hopping rates in \cite{Sasamoto}, \cite{barraquand} and \cite{Povolotsky}. Further we introduced integrable boundary conditions that were derived from the boundary Yang-Baxter equation using a general solution for the off-diagonal K-matrix. We studied the duality properties of this chain and exemplified how correlation functions in the stationary state can be computed. In the final section we show that the fluctuating hydrodynamics of the system without boundaries corresponds to the semiclassical evolution of a spinning string which naturally arises from the AdS/CFT dictionary.

The identification of the stochastic process with the non-compact spin chain allows to study the Markov process using powerful integrability tools. In particular one may expect that the steady state of the boundary model can be obtained from some sort of Bethe ansatz, see also \cite{Crampee} where the relation between the QISM approach and the matrix product ansatz has been discussed. Duality has told us that instead of computing the steady state we can compute the absorption probabilities of the dual process. This seems very similar to the Bethe ansatz which we have shown is equivalent to the one of  a spin chain with triangular K-matrices. The precise relation for this apparent connection between duality and Bethe ansatz deserves some further investigation.

The fact that we are obtaining quantum systems that yield, in imaginary times, the generators of stochastic dynamics,
may partially be argued from the fact that the spin chains obtained from $\mathcal{N}=4$ super Yang-Mills theory are naturally the bosonic parts of some larger supersymmetric
Hamiltonian. Indeed, in order to be stochastic and to have a stationary limit, the generator has to have a positive semi-definite spectrum, a fact that comes naturally from 
supersymmetry. This same supersymmetry guarantees that the equilibrium theorems for the dynamics of a system in contact with a stochastic bath hold.
The only thing that in the end one needs to check is the construction of one {\em or more} bases in which the matrix elements of the Hamiltonian are negative on the diagonal
and positive elsewhere, so that the probability rates are positive. 
Given what we have said, the temptation is great to conjecture that the bosonic part of $\mathcal{N}=4$ super Yang-Mills theory may be mapped (at least in the planar approximation) 
to a stochastic system, 
and that the fermions come from ``stochastic quantization''. However, currently it is not clear how and if the fundamental fields can be interpreted as the particles of a process or whether such particle picture can be found at all for the representation relevant.

We expect that the models studied in this article lift to the trigonometric/asymmetric case. Here the hopping rates of the stochastic process in the bulk are the ones of a q-Hahn antisymmetric process \cite{barraquand} and can be identified with non-compact  XXZ spin chain \cite{frassek00}. The spin $\frac{1}{2}$ case then yields the MADM \cite{Sasamoto}. Further, we suspect that the multi-species models, which were discussed in \cite{Kuniba}, arise from spin chains with symmetric representations of $\mathfrak{su}(r,1)$ and the corresponding q-deformation.  It would be interesting to derive the K-matrix in these cases and study the processes with reservoirs. The string picture of the trigonometric case with $r=1$ may be connected to a deformation of $AdS_5\times S^5$. 

Finally it would be interesting to study the relation to the stochastic R-matrix approach in \cite{Kuniba} and the stochastic K-matrix that has been recently obtained in \cite{Mangazeev:2019rzf} in connection to Baxter Q-operators as discussed at the end of Section~\ref{povo}.

\section*{Acknowledgments}
We like to thank Ivan Corwin, Gregory Korchemsky, Vivien Lecomte, Kirone Mallick and Rodrigo A. Pimenta for useful discussions. Further, we thank the anonymous referees for their useful comments.
R.F. likes to thank I.M. Szecsenyi for interesting discussions and collaboration on a related topic.
J.K. is   supported by the Simons Foundation  Grant No 454943.
R.F. was supported by the visitor program of the IHÉS where a significant part of this work has been carried out. 
R.F. also likes to thank the University of Modena and Reggio Emilia for hospitality. Finally we thank the organizers of RAQIS'18  at  LAPTh where this work originated.

\newpage
\appendix

\addtocontents{toc}{\protect\setcounter{tocdepth}{1}}

\section{Taylor expansion}
\label{appA}

\subsection{Derivation \eqref{levy:bulk}}
The action of the bulk generator $-\mathcal{H}^t_{i,i+1}$ 
on a function of the rescaled variables $(x_i^{(M)}, x_{i+1}^{(M)})$ reads
\begin{eqnarray}
-\mathcal{H}^t_{i,i+1} f(x_i^{(M)}, x_{i+1}^{(M)})
& = & 
\sum_{k=1}^{x_i M} \; \frac{1}{k} \Big[ f\Big(x_i^{(M)} - \frac{k}{M}, x_{i+1}^{(M)} + \frac{k}{M}\Big)   - f\Big(x_i^{(M)} , x_{i+1}^{(M)}\Big)  \Big] 
\nonumber\\
& + &
\sum_{k=1}^{x_i M} \; \frac{1}{k} \Big[ f\Big(x_i^{(M)} + \frac{k}{M}, x_{i+1}^{(M)} - \frac{k}{M}\Big)   - f\Big(x_i^{(M)} , x_{i+1}^{(M)}\Big)  \Big]. 
\end{eqnarray}
We only consider the scaling limit of the right jumps, the proof is analogous for the
left jumps. On one hand we notice that, by a bivariate Taylor expansion, we may write
\begin{eqnarray}
&&
\sum_{k=1}^{x_i M} \; \frac{1}{k} \Big[ f\Big(x_i^{(M)} - \frac{k}{M}, x_{i+1}^{(M)} + \frac{k}{M}\Big)   - f\Big(x_i^{(M)} , x_{i+1}^{(M)}\Big)  \Big]  =
\nonumber\\
&& 
\sum_{k=1}^{x_i M} \; \frac{1}{k} \sum_{a=1}^{\infty} \frac{1}{a!} \sum_{b=0}^a {a \choose b} 
\Big[\Big(\frac{\partial}{\partial x_i}\Big)^b \Big( \frac{\partial}{\partial x_{i+1}} \Big)^{a-b} 
f\Big(x_i^{(M)} , x_{i+1}^{(M)}\Big)\Big]
\Big(-\frac{k}{M} \Big)^b \Big( \frac{k}{M} \Big)^{a-b}.
\end{eqnarray}
This implies
\begin{eqnarray}
\label{hi1}
&& \lim_{M\to\infty}
\sum_{k=1}^{x_i M} \; \frac{1}{k} \Big[ f\Big(x_i^{(M)} - \frac{k}{M}, x_{i+1}^{(M)} + \frac{k}{M}\Big)   - f\Big(x_i^{(M)} , x_{i+1}^{(M)}\Big)  \Big]  =
\nonumber\\
&&
\sum_{a=1}^{\infty} \frac{1}{a!} \sum_{b=0}^a {a \choose b} 
\Big[\Big(\frac{\partial}{\partial x_i}\Big)^b \Big( \frac{\partial}{\partial x_{i+1}} \Big)^{a-b} 
f(x_i , x_{i+1})\Big] (-1)^b \int_{0}^{x_i} \alpha^{a-1} d\alpha
\end{eqnarray}
where the convergence of the Riemann sum 
\be
\label{conv1}
\sum_{k=1}^{x_iM} \frac{1}{k} \Big(\frac{k}{M} \Big)^a  \longrightarrow \int_{0}^{x_i} \alpha^{a-1} d\alpha  \qquad \text{as} \quad M\to\infty
\ee
has been exploited. 
On the other hand we may also write 
\begin{eqnarray}
\label{hi2}
& & 
\int_{0}^{x_i} \frac{d \alpha}{\alpha}\Big\{ f(x_i - \alpha, x_{i+1} + \alpha) - f(x_i,x_{i+1})\Big\} =
\nonumber\\
& &
\int_{0}^{x_i} \frac{d \alpha}{\alpha} \sum_{a=1}^{\infty} \frac{1}{a!} \sum_{b=0}^a {a \choose b} 
\Big(\frac{\partial}{\partial x_i}\Big)^b \Big( \frac{\partial}{\partial x_{i+1}} \Big)^{a-b} 
f(x_i, x_{i+1}) (-\alpha)^b (\alpha)^{a-b} =
\nonumber\\
& &
\int_{0}^{x_i} \frac{d \alpha}{\alpha} \sum_{a=1}^{\infty} \frac{1}{a!} \sum_{b=0}^a {a \choose b} 
\Big(\frac{\partial}{\partial x_i}\Big)^b \Big( \frac{\partial}{\partial x_{i+1}} \Big)^{a-b} 
f(x_i, x_{i+1}) (-1)^b \alpha^{a}.
\end{eqnarray}
The right hand sides of \eqref{hi1} and \eqref{hi2} do coincide, thus concluding the proof.

\subsection{Derivation \eqref{levy:boundary}}
The action of the bulk generator $-\mathcal{H}^t_{1}$ 
on a function of the rescaled variables $x_1^{(M)}$ reads
\begin{eqnarray}
-\mathcal{H}^t_{1} f(x_1^{(M)})
& = & 
\sum_{k=1}^{x_1 M} \; \frac{1}{k} \Big[f\Big(x_1^{(M)} - \frac{k}{M}\Big)   - f\Big(x_1^{(M)}\Big)  \Big] 
\nonumber\\
& + &
\sum_{k=1}^{\infty} \; \frac{[\beta_1^{(M)}]^k}{k} \Big[f\Big(x_1^{(M)} + \frac{k}{M}\Big)   - f\Big(x_1^{(M)}\Big)  \Big] . 
\end{eqnarray}
By Taylor expansion, we may write
\begin{eqnarray}
-\mathcal{H}^t_{1} f(x_1^{(M)})
& = & 
\sum_{k=1}^{x_1 M} \; \frac{1}{k} \sum_{n=1}^\infty \frac{1}{n!} f^{(n)}\Big(x_1^{(M)}\Big) \Big(-\frac{k}{M}\Big)^n  
\nonumber\\
& + &
\sum_{k=1}^{\infty} \; \frac{[\beta_1^{(M)}]^k}{k} \sum_{n=1}^\infty \frac{1}{n!} f^{(n)}\Big(x_1^{(M)}\Big) \Big(\frac{k}{M}\Big)^n   . 
\end{eqnarray}
This yields in the limit
\begin{eqnarray}
\label{hi3}
- \lim_{M\to\infty} \mathcal{H}^t_{1} f(x_1^{(M)})
& = & 
\sum_{n=1}^\infty \frac{1}{n!} f^{(n)}(x_1) (-1)^n \int_{0}^{x_1} \alpha^{n-1} d\alpha
\nonumber\\
& + &
\sum_{n=1}^\infty \frac{1}{n!} f^{(n)}(x_1) \int_{0}^{\infty} \alpha^{n-1} e^{-\lambda_1 \alpha} d\alpha.
\end{eqnarray}
where, besides \eqref{conv1}, it has been used the convergence  
\be
\sum_{k=1}^{\infty}\frac{[\beta_1^{(M)}]^k}{k} \Big(\frac{k}{M} \Big)^n
=
\sum_{k=1}^{\infty}\frac{[1-\frac{\lambda_1}{M}]^k}{k} \Big(\frac{k}{M} \Big)^n 
 \longrightarrow \int_{0}^{\infty} \alpha^{n-1} e^{-\lambda_1 \alpha} d\alpha,  \qquad \text{as} \quad M\to\infty.
\ee
Clearly, the right hand sides of \eqref{hi3} coincides with the right hand side of \eqref{levy:boundary}.

\section{Infinite sums}

\subsection{Derivation \eqref{eq:coef1}}\label{ap:comp1}
It is straightforward to see that \eqref{eq:coef1} is true for $k\leq l$. In the case $k>l$ the sum reduces to 
\begin{equation}
  \sum_{m=l}^k \binom{k}{m}\binom{m}{l}\beta^{k-m}(-\beta)^{m-l}\psi(m+1)=\sum_{m=l}^k\sum_{r=1}^m\binom{k}{m}\binom{m}{l}\frac{(-1)^{m-l}\beta^{k-l}}{m+1-r}=-\frac{\beta^{k-l}}{k-l}
\end{equation} 
where we first used
\begin{equation}
 \psi(m+1)=\psi(1)+\sum_{r=1}^m\frac{1}{m+1-r}
\end{equation} 
and exchanged the sums in the second step.

\subsection{Derivation \eqref{eq:coef2}}\label{ap:comp2}
After noting that the first sum truncates and shifting the second sum we rewrite \eqref{eq:coef2} as
\begin{equation}
\begin{split}
 \sum_{m_2=1}^\infty\frac{(-1)^{l+m_2} }{m_2}\alpha^{m_2}\beta ^{k-l+m_2}& \binom{m_2}{l} \, _2F_1(-k,m_2+1;-l+m_2+1;1) =\qquad \qquad\qquad\\
 &=  \sum_{m_2=1}^\infty(-1)^{l+m_2}\alpha^{m_2}\beta ^{k-l+m_2} \frac{\Gamma(m_2)\Gamma (k-l)}{\Gamma(l+1)\Gamma (-l) \Gamma (k-l+m_2+1)}\\
 \end{split}
\end{equation} 
The case $k>l$ in \eqref{eq:coef2}  is straightforward, the denominator is finite while the numerator diverges. For $k\leq l$ one has to be more careful. 
One gets
\begin{equation}
\begin{split}
\sum_{m_2=1}^\infty(-1)^{l+m_2}\alpha^{m_2}\beta ^{k-l+m_2}&\frac{\Gamma(m_2)\Gamma (k-l)}{\Gamma(l+1)\Gamma (-l) \Gamma (k-l+m_2+1)}\\
 &=\frac{(-1)^{k+l}}{(l-k)!}\sum_{m_2=1}^\infty(-1)^{m_2}\alpha^{m_2}\beta ^{k-l+m_2} \frac{\Gamma(m_2)}{\Gamma (k-l+m_2+1)}\\
 \end{split}
\end{equation} 
To show the remaining relations we substitute 
\begin{equation}
 \alpha=\frac{1}{1-\beta}=\sum_{p=0}^\infty \beta^p
\end{equation} 
Using the binomial series 
\begin{equation}\label{eq:binomial}
\alpha^{m_2}= \frac{1}{(1-\beta)^{m_2}}=\sum_{n=0}^\infty\frac{\Gamma(m_2+n)}{\Gamma(m_2)\Gamma(1+n)}\beta^n
\end{equation} 
we get for $k=l$
\begin{equation}
\begin{split}
\sum_{m_2=1}^\infty\frac{(-1)^{m_2} \alpha^{m_2} \beta^{m_2}}{m_2}&=\sum_{n=0}^\infty\sum_{m_2=1}^\infty\frac{(-1)^{m_2} \beta^{n+m_2}(m_2+n-1)!}{n!m_2!}\\
&=\sum_{q=1}^\infty\sum_{m_2=1}^q\frac{(-1)^{m_2} \beta^{q}(q-1)!}{(q-m_2)!m_2!}=-\sum_{q=1}^\infty\frac{\beta^{q}}{q}\\
\end{split}
\end{equation} 
For $k<l$ we get 
\begin{equation}
\begin{split}
 &\quad\,\,\frac{(-1)^{k+l}\beta^{k-l}}{(l-k)!}\sum_{m_2=l-k}^\infty\sum_{n=0}^\infty\frac{(-1)^{m_2} \beta^{n+m_2}(m_2+n-1)!}{n!(k-l+m_2)!}\\&
 = \frac{(-1)^{k+l}\beta^{k-l}}{(l-k)!}\sum_{m_2=l-k}^\infty\sum_{q=m_2}^\infty\frac{(-1)^{m_2} \beta^{q}(q-1)!}{(q-m_2)!(k-l+m_2)!}
 \\ &= \frac{(-1)^{k+l}\beta^{k-l}}{(l-k)!}\sum_{q=l-k}^\infty\sum_{m_2=l-k}^q\frac{(-1)^{m_2} \beta^{q}(q-1)!}{(q-m_2)!(k-l+m_2)!}
 \end{split}
\end{equation} 
Finally noting that 
\begin{equation}
 \sum_{m=0}^p\frac{(-1)^{m}}{(p-m)!m!}=\delta_{p,0}
\end{equation} 
we obtain the final result.

{
\small
\bibliographystyle{utphys2}
\bibliography{refs}

\providecommand{\href}[2]{#2}\begingroup\raggedright\begin{thebibliography}{10}

\bibitem{Sasamoto}
T.~Sasamoto and M.~Wadati, ``One-dimensional asymmetric diffusion model without
  exclusion,'' \href{http://dx.doi.org/10.1103/PhysRevE.58.4181}{{\em Phys.
  Rev. E} {\bfseries 58} (Oct, 1998) 4181--4190}.

\bibitem{barraquand}
G.~Barraquand and I.~Corwin, ``The $q$-Hahn asymmetric exclusion process,''
  \href{http://dx.doi.org/10.1214/15-AAP1148}{{\em Ann. Appl. Probab.}
  {\bfseries 26} no.~4, (08, 2016) 2304--2356},
  \href{http://arxiv.org/abs/1501.03445}{{\ttfamily arXiv:1501.03445
  [math.PR]}}.

\bibitem{Povolotsky}
A.~M. Povolotsky, ``On the integrability of zero-range chipping models with
  factorized steady states,''
  \href{http://dx.doi.org/10.1088/1751-8113/46/46/465205}{{\em Journal of
  Physics A: Mathematical and Theoretical} {\bfseries 46} no.~46, (Nov, 2013)
  465205}, \href{http://arxiv.org/abs/1308.3250}{{\ttfamily arXiv:1308.3250
  [math-ph]}}.

\bibitem{kipnis1982heat}
C.~Kipnis, C.~Marchioro, and E.~Presutti, ``Heat flow in an exactly solvable
  model,'' {\em Journal of Statistical Physics} {\bfseries 27} no.~1, (1982)
  65--74.

\bibitem{giardina2007duality}
C.~Giardina, J.~Kurchan, and F.~Redig, ``Duality and exact correlations for a
  model of heat conduction,'' {\em Journal of mathematical physics} {\bfseries
  48} no.~3, (2007) 033301,
  \href{http://arxiv.org/abs/cond-mat/0612198}{{\ttfamily
  arXiv:cond-mat/0612198 [cond-mat]}}.

\bibitem{GKR}
C.~Giardin{\`a}, J.~Kurchan, F.~Redig, and K.~Vafayi, ``Duality and Hidden
  Symmetries in Interacting Particle Systems,''
  \href{http://dx.doi.org/10.1007/s10955-009-9716-2}{{\em Journal of
  Statistical Physics} {\bfseries 135} no.~1, (Apr, 2009) 25--55},
  \href{http://arxiv.org/abs/0810.1202}{{\ttfamily arXiv:0810.1202 [math-ph]}}.

\bibitem{Bethe1931}
H.~Bethe, ``Zur Theorie der Metalle,''
  \href{http://dx.doi.org/10.1007/BF01341708}{{\em Zeitschrift f{\"u}r Physik}
  {\bfseries 71} no.~3, (Mar, 1931) 205--226}.

\bibitem{Lipatov:1993yb}
L.~N. Lipatov, ``{Asymptotic behavior of multicolor QCD at high energies in
  connection with exactly solvable spin models},'' {\em JETP Lett.} {\bfseries
  59} (1994) 596--599, \href{http://arxiv.org/abs/hep-th/9311037}{{\ttfamily
  arXiv:hep-th/9311037 [hep-th]}}.
[Pisma Zh. Eksp. Teor. Fiz.59,571(1994)].

\bibitem{Faddeev:1994zg}
L.~D. Faddeev and G.~P. Korchemsky, ``{High-energy QCD as a completely
  integrable model},''
  \href{http://dx.doi.org/10.1016/0370-2693(94)01363-H}{{\em Phys. Lett.}
  {\bfseries B342} (1995) 311--322},
\href{http://arxiv.org/abs/hep-th/9404173}{{\ttfamily arXiv:hep-th/9404173
  [hep-th]}}.

\bibitem{Braun:1998id}
V.~M. Braun, S.~E. Derkachov, and A.~N. Manashov, ``{Integrability of three
  particle evolution equations in QCD},''
  \href{http://dx.doi.org/10.1103/PhysRevLett.81.2020}{{\em Phys. Rev. Lett.}
  {\bfseries 81} (1998) 2020--2023},
\href{http://arxiv.org/abs/hep-ph/9805225}{{\ttfamily arXiv:hep-ph/9805225
  [hep-ph]}}.

\bibitem{Minahan:2002ve}
J.~A. Minahan and K.~Zarembo, ``{The Bethe ansatz for N=4 superYang-Mills},''
  \href{http://dx.doi.org/10.1088/1126-6708/2003/03/013}{{\em JHEP} {\bfseries
  03} (2003) 013},
\href{http://arxiv.org/abs/hep-th/0212208}{{\ttfamily arXiv:hep-th/0212208
  [hep-th]}}.

\bibitem{Beisert:2003yb}
N.~Beisert and M.~Staudacher, ``{The N=4 SYM integrable super spin chain},''
  \href{http://dx.doi.org/10.1016/j.nuclphysb.2003.08.015}{{\em Nucl. Phys.}
  {\bfseries B670} (2003) 439--463},
\href{http://arxiv.org/abs/hep-th/0307042}{{\ttfamily arXiv:hep-th/0307042
  [hep-th]}}.

\bibitem{Beisert:2003jj}
N.~Beisert, ``{The complete one loop dilatation operator of N=4 superYang-Mills
  theory},'' \href{http://dx.doi.org/10.1016/j.nuclphysb.2003.10.019}{{\em
  Nucl. Phys.} {\bfseries B676} (2004) 3--42},
\href{http://arxiv.org/abs/hep-th/0307015}{{\ttfamily arXiv:hep-th/0307015
  [hep-th]}}.

\bibitem{kruczenski}
M.~Kruczenski, ``{Spin chains and string theory},''
  \href{http://dx.doi.org/10.1103/PhysRevLett.93.161602}{{\em Phys. Rev. Lett.}
  {\bfseries 93} (2004) 161602},
\href{http://arxiv.org/abs/hep-th/0311203}{{\ttfamily arXiv:hep-th/0311203
  [hep-th]}}.

\bibitem{bellucci-sl2}
S.~Bellucci, P.~Y. Casteill, J.~F. Morales, and C.~Sochichiu, ``{SL(2) spin
  chain and spinning strings on AdS(5) x S**5},''
  \href{http://dx.doi.org/10.1016/j.nuclphysb.2004.11.020}{{\em Nucl. Phys.}
  {\bfseries B707} (2005) 303--320},
\href{http://arxiv.org/abs/hep-th/0409086}{{\ttfamily arXiv:hep-th/0409086
  [hep-th]}}.

\bibitem{Stefanski:2004cw}
B.~Stefanski, Jr. and A.~A. Tseytlin, ``{Large spin limits of AdS/CFT and
  generalized Landau-Lifshitz equations},''
  \href{http://dx.doi.org/10.1088/1126-6708/2004/05/042}{{\em JHEP} {\bfseries
  05} (2004) 042},
\href{http://arxiv.org/abs/hep-th/0404133}{{\ttfamily arXiv:hep-th/0404133
  [hep-th]}}.

\bibitem{kardar1986dynamic}
M.~Kardar, G.~Parisi, and Y.-C. Zhang, ``Dynamic scaling of growing
  interfaces,'' {\em Physical Review Letters} {\bfseries 56} no.~9, (1986) 889.

\bibitem{sasamoto2010one}
T.~Sasamoto and H.~Spohn, ``One-dimensional Kardar-Parisi-Zhang equation: an
  exact solution and its universality,'' {\em Physical review letters}
  {\bfseries 104} no.~23, (2010) 230602,
  \href{http://arxiv.org/abs/1002.1883}{{\ttfamily arXiv:1002.1883
  [cond-mat]}}.

\bibitem{corwin2012kardar}
I.~Corwin, ``The Kardar--Parisi--Zhang equation and universality class,'' {\em
  Random matrices: Theory and applications} {\bfseries 1} no.~01, (2012)
  1130001, \href{http://arxiv.org/abs/1106.1596}{{\ttfamily arXiv:1106.1596
  [math.PR]}}.

\bibitem{alimohammadi1998exact}
M.~Alimohammadi, V.~Karimipour, and M.~Khorrami, ``Exact solution of a
  one-parameter family of asymmetric exclusion processes,'' {\em Physical
  Review E} {\bfseries 57} no.~6, (1998) 6370.

\bibitem{alimohammadi1999two}
M.~Alimohammadi, V.~Karimipour, and M.~Khorrami, ``A two-parametric family of
  asymmetric exclusion processes and its exact solution,'' {\em Journal of
  statistical physics} {\bfseries 97} no.~1-2, (1999) 373--394.

\bibitem{Derkachov:1999pz}
S.~E. Derkachov, ``{Baxter's Q-operator for the homogeneous XXX spin chain},''
  \href{http://dx.doi.org/10.1088/0305-4470/32/28/309}{{\em J. Phys.}
  {\bfseries A32} (1999) 5299--5316},
\href{http://arxiv.org/abs/solv-int/9902015}{{\ttfamily arXiv:solv-int/9902015
  [solv-int]}}.

\bibitem{Kuniba}
A.~Kuniba, V.~V. Mangazeev, S.~Maruyama, and M.~Okado, ``{Stochastic R matrix
  for $U_q(A_n^{(1)})$},''
  \href{http://dx.doi.org/10.1016/j.nuclphysb.2016.09.016}{{\em Nucl. Phys.}
  {\bfseries B913} (2016) 248--277},
\href{http://arxiv.org/abs/1604.08304}{{\ttfamily arXiv:1604.08304 [math.QA]}}.

\bibitem{barraquand2017random}
G.~Barraquand and I.~Corwin, ``Random-walk in Beta-distributed random
  environment,'' {\em Probability Theory and Related Fields} {\bfseries 167}
  no.~3-4, (2017) 1057--1116, \href{http://arxiv.org/abs/1503.04117}{{\ttfamily
  arXiv:1503.04117 [math.PR]}}.

\bibitem{thiery2015integrable}
T.~Thiery and P.~Le~Doussal, ``On integrable directed polymer models on the
  square lattice,'' {\em Journal of Physics A: Mathematical and Theoretical}
  {\bfseries 48} no.~46, (2015) 465001,
  \href{http://arxiv.org/abs/1506.05006}{{\ttfamily arXiv:1506.05006
  [cond-mat.dis-nn]}}.

\bibitem{Sklyanin:1988yz}
E.~K. Sklyanin, ``{Boundary Conditions for Integrable Quantum Systems},''
\href{http://dx.doi.org/10.1088/0305-4470/21/10/015}{{\em J. Phys.} {\bfseries
  A21} (1988) 2375--289}.

\bibitem{Tailleur}
J.~Tailleur, J.~Kurchan, and V.~Lecomte, ``Mapping out-of-equilibrium into
  equilibrium in one-dimensional transport models,''
  \href{http://dx.doi.org/10.1088/1751-8113/41/50/505001}{{\em Journal of
  Physics A: Mathematical and Theoretical} {\bfseries 41} no.~50, (Nov, 2008)
  505001}, \href{http://arxiv.org/abs/0809.0709}{{\ttfamily arXiv:0809.0709
  [cond-mat]}}.

\bibitem{jona}
L.~Bertini, A.~De~Sole, D.~Gabrielli, G.~Jona-Lasinio, and C.~Landim,
  ``Macroscopic fluctuation theory for stationary non-equilibrium states,''
  {\em Journal of Statistical Physics} {\bfseries 107} no.~3-4, (2002)
  635--675, \href{http://arxiv.org/abs/cond-mat/0108040}{{\ttfamily
  arXiv:cond-mat/0108040 [cond-mat]}}.

\bibitem{Faddeev:1996iy}
L.~D. Faddeev, ``{How algebraic Bethe ansatz works for integrable model},'' in
  {\em {Relativistic gravitation and gravitational radiation. Proceedings,
  School of Physics, Les Houches, France, September 26-October 6, 1995}},
  pp.~pp. 149--219.
\newblock 1996.
\newblock
\href{http://arxiv.org/abs/hep-th/9605187}{{\ttfamily arXiv:hep-th/9605187
  [hep-th]}}.
\newblock

\bibitem{Korchemsky:1994um}
G.~P. Korchemsky, ``{Bethe ansatz for QCD pomeron},''
  \href{http://dx.doi.org/10.1016/0550-3213(95)00099-E}{{\em Nucl. Phys.}
  {\bfseries B443} (1995) 255--304},
\href{http://arxiv.org/abs/hep-ph/9501232}{{\ttfamily arXiv:hep-ph/9501232
  [hep-ph]}}.

\bibitem{frassek00}
R.~Frassek, ``{The non-compact XXZ spin chain as stochastic particle
  process},'' \href{http://dx.doi.org/10.1088/1751-8121/ab2fb1}{{\em J. Phys.}
  {\bfseries A52} no.~33, (2019) 335202},
\href{http://arxiv.org/abs/1904.02191}{{\ttfamily arXiv:1904.02191 [math-ph]}}.

\bibitem{Martins:2009dt}
M.~J. Martins and C.~S. Melo, ``{Algebraic Bethe ansatz for U(1) Invariant
  Integrable Models: Compact and non-Compact Applications},''
  \href{http://dx.doi.org/10.1016/j.nuclphysb.2009.04.018}{{\em Nucl. Phys.}
  {\bfseries B820} (2009) 620--648},
\href{http://arxiv.org/abs/0902.3476}{{\ttfamily arXiv:0902.3476 [math-ph]}}.

\bibitem{Frassek:2012mg}
R.~Frassek and C.~Meneghelli, ``{From Baxter Q-Operators to Local Charges},''
  \href{http://dx.doi.org/10.1088/1742-5468/2013/02/P02019}{{\em J. Stat.
  Mech.} {\bfseries 1302} (2013) P02019},
\href{http://arxiv.org/abs/1207.4513}{{\ttfamily arXiv:1207.4513 [hep-th]}}.

\bibitem{Bazhanov:2010ts}
V.~V. Bazhanov, T.~Lukowski, C.~Meneghelli, and M.~Staudacher, ``{A Shortcut to
  the Q-Operator},''
  \href{http://dx.doi.org/10.1088/1742-5468/2010/11/P11002}{{\em J. Stat.
  Mech.} {\bfseries 1011} (2010) P11002},
\href{http://arxiv.org/abs/1005.3261}{{\ttfamily arXiv:1005.3261 [hep-th]}}.

\bibitem{Bazhanov:2010jq}
V.~V. Bazhanov, R.~Frassek, T.~Lukowski, C.~Meneghelli, and M.~Staudacher,
  ``{Baxter Q-Operators and Representations of Yangians},''
  \href{http://dx.doi.org/10.1016/j.nuclphysb.2011.04.006}{{\em Nucl. Phys.}
  {\bfseries B850} (2011) 148--174},
\href{http://arxiv.org/abs/1010.3699}{{\ttfamily arXiv:1010.3699 [math-ph]}}.

\bibitem{Frassek:2011aa}
R.~Frassek, T.~Lukowski, C.~Meneghelli, and M.~Staudacher, ``{Baxter Operators
  and Hamiltonians for 'nearly all' Integrable Closed $\mathfrak{gl}(n)$ Spin
  Chains},'' \href{http://dx.doi.org/10.1016/j.nuclphysb.2013.06.006}{{\em
  Nucl. Phys.} {\bfseries B874} (2013) 620--646},
\href{http://arxiv.org/abs/1112.3600}{{\ttfamily arXiv:1112.3600 [math-ph]}}.

\bibitem{Frassek:2010ga}
R.~Frassek, T.~Lukowski, C.~Meneghelli, and M.~Staudacher, ``{Oscillator
  Construction of su(n|m) Q-Operators},''
  \href{http://dx.doi.org/10.1016/j.nuclphysb.2011.04.008}{{\em Nucl. Phys.}
  {\bfseries B850} (2011) 175--198},
\href{http://arxiv.org/abs/1012.6021}{{\ttfamily arXiv:1012.6021 [math-ph]}}.

\bibitem{Frassek:2017bfz}
R.~Frassek, C.~Marboe, and D.~Meidinger, ``{Evaluation of the operatorial
  Q-system for non-compact super spin chains},''
  \href{http://dx.doi.org/10.1007/JHEP09(2017)018}{{\em JHEP} {\bfseries 09}
  (2017) 018},
\href{http://arxiv.org/abs/1706.02320}{{\ttfamily arXiv:1706.02320 [hep-th]}}.

\bibitem{Derkachov:1999ze}
S.~E. Derkachov, G.~P. Korchemsky, and A.~N. Manashov, ``{Evolution equations
  for quark gluon distributions in multicolor QCD and open spin chains},''
  \href{http://dx.doi.org/10.1016/S0550-3213(99)00702-6}{{\em Nucl. Phys.}
  {\bfseries B566} (2000) 203--251},
\href{http://arxiv.org/abs/hep-ph/9909539}{{\ttfamily arXiv:hep-ph/9909539
  [hep-ph]}}.

\bibitem{Derkachov:2003qb}
S.~E. Derkachov, G.~P. Korchemsky, and A.~N. Manashov, ``{Baxter Q operator and
  separation of variables for the open SL(2,R) spin chain},''
  \href{http://dx.doi.org/10.1088/1126-6708/2003/10/053}{{\em JHEP} {\bfseries
  10} (2003) 053},
\href{http://arxiv.org/abs/hep-th/0309144}{{\ttfamily arXiv:hep-th/0309144
  [hep-th]}}.

\bibitem{Belitsky:2014rba}
A.~V. Belitsky, S.~E. Derkachov, and A.~N. Manashov, ``{Quantum mechanics of
  null polygonal Wilson loops},''
  \href{http://dx.doi.org/10.1016/j.nuclphysb.2014.03.007}{{\em Nucl. Phys.}
  {\bfseries B882} (2014) 303--351},
\href{http://arxiv.org/abs/1401.7307}{{\ttfamily arXiv:1401.7307 [hep-th]}}.

\bibitem{deVega:1992zd}
H.~J. de~Vega and A.~Gonzalez~Ruiz, ``{Boundary K matrices for the six vertex
  and the n(2n-1) A(n-1) vertex models},''
  \href{http://dx.doi.org/10.1088/0305-4470/26/12/007}{{\em J. Phys.}
  {\bfseries A26} (1993) L519--L524},
\href{http://arxiv.org/abs/hep-th/9211114}{{\ttfamily arXiv:hep-th/9211114
  [hep-th]}}.

\bibitem{Kulish:1981gi}
P.~P. Kulish, N.~{\relax Yu}. Reshetikhin, and E.~K. Sklyanin, ``{Yang-Baxter
  Equation and Representation Theory. 1.},''
\href{http://dx.doi.org/10.1007/BF02285311}{{\em Lett. Math. Phys.} {\bfseries
  5} (1981) 393--403}.

\bibitem{Baseilhac:2017hoz}
P.~Baseilhac and Z.~Tsuboi, ``{Asymptotic representations of augmented
  q-Onsager algebra and boundary K-operators related to Baxter Q-operators},''
  \href{http://dx.doi.org/10.1016/j.nuclphysb.2018.02.017}{{\em Nucl. Phys.}
  {\bfseries B929} (2018) 397--437},
\href{http://arxiv.org/abs/1707.04574}{{\ttfamily arXiv:1707.04574 [math-ph]}}.

\bibitem{Frassek:2015mra}
R.~Frassek and I.~M. Szecsenyi, ``{Q-operators for the open Heisenberg spin
  chain},'' \href{http://dx.doi.org/10.1016/j.nuclphysb.2015.10.010}{{\em Nucl.
  Phys.} {\bfseries B901} (2015) 229--248},
\href{http://arxiv.org/abs/1509.04867}{{\ttfamily arXiv:1509.04867 [math-ph]}}.

\bibitem{Braun:2018fiz}
V.~M. Braun, Y.~Ji, and A.~N. Manashov, ``{Integrability in heavy quark
  effective theory},'' \href{http://dx.doi.org/10.1007/JHEP06(2018)017}{{\em
  JHEP} {\bfseries 06} (2018) 017},
\href{http://arxiv.org/abs/1804.06289}{{\ttfamily arXiv:1804.06289 [hep-th]}}.

\bibitem{Belitsky:2019ygi}
A.~V. Belitsky, ``{Separation of Variables for a flux tube with an end},''
\href{http://arxiv.org/abs/1902.08596}{{\ttfamily arXiv:1902.08596 [hep-th]}}.

\bibitem{Belliard2013}
S.~Belliard, N.~Cramp{\'e}, and E.~Ragoucy, ``Algebraic Bethe Ansatz for Open
  XXX Model with Triangular Boundary Matrices,''
  \href{http://dx.doi.org/10.1007/s11005-012-0601-6}{{\em Letters in
  Mathematical Physics} {\bfseries 103} no.~5, (May, 2013) 493--506},
  \href{http://arxiv.org/abs/1209.4269}{{\ttfamily arXiv:1209.4269 [math-ph]}}.

\bibitem{Antonio:2014qxa}
N.~Cirilo~António, N.~Manojlović, and I.~Salom, ``{Algebraic Bethe ansatz for
  the XXX chain with triangular boundaries and Gaudin model},''
  \href{http://dx.doi.org/10.1016/j.nuclphysb.2014.10.014}{{\em Nucl. Phys.}
  {\bfseries B889} (2014) 87--108},
\href{http://arxiv.org/abs/1405.7398}{{\ttfamily arXiv:1405.7398 [math-ph]}}.

\bibitem{franceschini2017stochastic}
C.~Franceschini and C.~Giardina, ``Stochastic duality and orthogonal
  polynomials,'' \href{http://arxiv.org/abs/1701.09115}{{\ttfamily
  arXiv:1701.09115 [math.PR]}}.

\bibitem{franceschini2018self}
C.~Franceschini, C.~Giardina, and W.~Groenevelt, ``Self-duality of Markov
  processes and intertwining functions,'' {\em Mathematical Physics, Analysis
  and Geometry} {\bfseries 21} no.~4, (2018) 29.

\bibitem{carinci2018orthogonal}
G.~Carinci, C.~Franceschini, C.~Giardina, W.~Groenevelt, and F.~Redig,
  ``Orthogonal dualities of Markov processes and unitary symmetries,''
  \href{http://arxiv.org/abs/1812.08553}{{\ttfamily arXiv:1812.08553
  [math.PR]}}.

\bibitem{Tseytlin:2004xa}
A.~A. Tseytlin, ``{Semiclassical strings and AdS/CFT},'' in {\em {String
  theory: From gauge interactions to cosmology. Proceedings, NATO Advanced
  Study Institute, Cargese, France, June 7-19, 2004}}, pp.~265--290.
\newblock 2004.
\newblock
\href{http://arxiv.org/abs/hep-th/0409296}{{\ttfamily arXiv:hep-th/0409296
  [hep-th]}}.
\newblock

\bibitem{Plefka:2005bk}
J.~Plefka, ``{Spinning strings and integrable spin chains in the AdS/CFT
  correspondence},'' \href{http://dx.doi.org/10.12942/lrr-2005-9}{{\em Living
  Rev. Rel.} {\bfseries 8} (2005) 9},
\href{http://arxiv.org/abs/hep-th/0507136}{{\ttfamily arXiv:hep-th/0507136
  [hep-th]}}.

\bibitem{Beisert:2010jr}
N.~Beisert {\em et~al.}, ``{Review of AdS/CFT Integrability: An Overview},''
  \href{http://dx.doi.org/10.1007/s11005-011-0529-2}{{\em Lett. Math. Phys.}
  {\bfseries 99} (2012) 3--32},
\href{http://arxiv.org/abs/1012.3982}{{\ttfamily arXiv:1012.3982 [hep-th]}}.

\bibitem{Zinn}
J.~Zinn-Justin, ``{Quantum field theory and critical phenomena},''
{\em Int. Ser. Monogr. Phys.} {\bfseries 113} (2002) 1--1054.

\bibitem{barci}
Z.~G. Arenas and D.~G. Barci, ``{Supersymmetric formulation of multiplicative
  white-noise stochastic processes},''
  \href{http://dx.doi.org/10.1103/PhysRevE.85.041122}{{\em Phys. Rev.}
  {\bfseries E85} (2012) 041122},
\href{http://arxiv.org/abs/1111.6123}{{\ttfamily arXiv:1111.6123
  [cond-mat.stat-mech]}}.

\bibitem{bellucci-super}
S.~Bellucci, P.~Y. Casteill, and J.~F. Morales, ``{Superstring sigma models
  from spin chains: The SU(1,1|1) case},''
  \href{http://dx.doi.org/10.1016/j.nuclphysb.2005.09.012}{{\em Nucl. Phys.}
  {\bfseries B729} (2005) 163--178},
\href{http://arxiv.org/abs/hep-th/0503159}{{\ttfamily arXiv:hep-th/0503159
  [hep-th]}}.

\bibitem{dijkgraaf}
R.~Dijkgraaf, D.~Orlando, and S.~Reffert, ``{Relating Field Theories via
  Stochastic Quantization},''
  \href{http://dx.doi.org/10.1016/j.nuclphysb.2009.07.018}{{\em Nucl. Phys.}
  {\bfseries B824} (2010) 365--386},
\href{http://arxiv.org/abs/0903.0732}{{\ttfamily arXiv:0903.0732 [hep-th]}}.

\bibitem{six}
J.~Kurchan, ``Six out of equilibrium lectures,'' {\em Lecture Notes of the Les
  Houches Summer School: Volume 90} (2009) ,
  \href{http://arxiv.org/abs/0901.1271}{{\ttfamily arXiv:0901.1271
  [cond-mat]}}.

\bibitem{Crampee}
N.~Crampe, E.~Ragoucy, and M.~Vanicat, ``{Integrable approach to simple
  exclusion processes with boundaries. Review and progress},''
  \href{http://dx.doi.org/10.1088/1742-5468/2014/11/P11032}{{\em J. Stat.
  Mech.} {\bfseries 1411} no.~11, (2014) P11032},
\href{http://arxiv.org/abs/1408.5357}{{\ttfamily arXiv:1408.5357 [math-ph]}}.

\bibitem{Mangazeev:2019rzf}
V.~V. Mangazeev and X.~Lu, ``{Boundary matrices for the higher spin six vertex
  model},'' \href{http://dx.doi.org/10.1016/j.nuclphysb.2019.114665}{{\em Nucl.
  Phys.} {\bfseries B945} (2019) 114665},
\href{http://arxiv.org/abs/1903.00274}{{\ttfamily arXiv:1903.00274 [math-ph]}}.

\end{thebibliography}\endgroup
}
\end{document}